\documentclass[intlimits,twoside,a4paper]{article}

\usepackage[cp1251]{inputenc}

\usepackage[eqsecnum]{cmpj3}

\usepackage{bm}

\DeclareMathOperator{\Tr}{Tr}
\DeclareMathOperator{\adj}{adj}

\allowdisplaybreaks

\issue{2020}{23}{1}{13703}
\doinumber{10.5488/CMP.23.13703}

\title{Thermoelectric properties of Mott insulator with correlated hopping at microdoping}
\author{D.A. Dobushovskyi, A.M. Shvaika}
\address{Institute for Condensed Matter Physics of the National Academy of Sciences of Ukraine, \\1 Svientsitskii St., 79011 Lviv, Ukraine
}

\date{Received August 16, 2019}

\begin{document}
	
	\maketitle
	
	\begin{abstract} 
		An influence of the localization of itinerant electrons induced by correlated hopping on the electronic charge and heat transport is discussed for the lightly doped Mott insulator phase of the Falicov-Kimball model. The case of strongly reduced hopping amplitude between the sites with occupied \(f\)-electron levels, when an additional band of localized \(d\)-electron states could appear on the DOS in the Mott gap, is considered. Due to the electron-hole asymmetry and anomalous features on the DOS and transport function induced by correlated hopping, a strong enhancement of the Seebeck coefficient is observed at low temperatures, when the flattened dependence is displayed in a wide temperature range.
		
	\end{abstract}

\section{Introduction}

During the last decade, various systems with specific electron properties have attracted a great attention of investigators. They include one- and two-dimensional organic conductors, three-dimensional solids and topological conductors, and right until the fermionic atoms on optical lattices. A great part of their properties can be explained exclusively by a proper treating of the electron dynamics involving the many-electron effects and electron correlations. One of the promising applications of such systems is the power generation or cooling by means of the thermoelectric effect. Theoretical description of the thermoelectric transport in strongly correlated electron systems is a challenge and requires the development of  new approaches, see, e.g., reference~\cite{costi:2519}, and most of the previous investigations were performed for the models with local single-site correlations of the Hubbard or Anderson type.

However, it was originally pointed by Hubbard~\cite{hubbard:238} that the second quantized representation of the inter-electron Coulomb interaction contains, besides the local term \(U\sum_i n_{i\uparrow}n_{i\downarrow}\), the nonlocal contributions including the inter-site Coulomb interaction \(\sum_{ij} V_{ij} \hat{n}_{i}\hat{n}_{j}\) and the so-called correlated hopping
\begin{equation}
\sum_{ij\sigma} t_{ij}^{(2)} (\hat{n}_{i\bar{\sigma}}+\hat{n}_{j\bar{\sigma}})c_{i\sigma}^{\dagger}c_{j\sigma}\,,
\end{equation}
which introduces new physical effects because now the value of inter-site hopping depends on the occupation of these states. 

Local Coulomb interaction is a subject of the famous Hubbard model and has been investigated for many decades in the theory of strongly correlated electron systems, whereas the correlated hopping attracts much less attention. Mainly, it was considered in connection with the elaboration of new mechanisms for high temperature superconductivity~\cite{hirsch:326,bulka:10303}, description of organic compounds~\cite{arrachea:1173} and molecular crystals~\cite{tsuchiizu:044519}, electron-hole asymmetry~\cite{didukh:7893}, and enhancement of magnetic properties~\cite{kollar:045107}.
The recent years show that correlated hopping is an important puzzle in the physics of quantum dots~\cite{meir:196802,hubsch:196401,tooski:055001} and it appears in a natural way in modelling the fermionic~\cite{jurgensen:043623,liberto:013624} and bosonic~\cite{eckholt:093028,luhmann:033021,jurgensen:193003} atoms on optical lattices.
However, due to its nonlocal character, the theoretical treatment of correlated hopping is difficult  and, in most cases, the solutions can only be obtained by rather drastic approximations.

The exact results, that can be obtained in some special cases, are of great importance, because they can be used for benchmarking various  approximations. 
In this article, we examine the Falicov-Kimball model~\cite{falicov:997}, the simplest model  of strongly correlated electrons, which considers the local interaction between the itinerant \(d\) electrons and localized \(f\) electrons.
It is a binary alloy type model and it displays a variety of modulated 
phases~\cite{gajek:3473,wojtkiewicz:233103,wojtkiewicz:3467,cencarikova:42701} in the ground state phase diagram for the one-dimensional \((D=1)\) 
and two-dimensional \((D=2)\) cases.
The main advantage of the Falicov-Kimball model is the featuring of an exact solution in infinite dimensions~\cite{brandt:365,freericks:1333} within the dynamical mean field theory (DMFT)~\cite{metzner:324,georges:13}.
Its extension by inclusion of correlated hopping was also considered and the DMFT solution with a nonlocal self-energy was provided~\cite{schiller:15660,shvaika:075101}.

In our previous articles~\cite{dobushovskyi:125133,dobushovskyi:23702}, we considered the charge and heat transport as well as optical conductivity spectra for the Falicov-Kimball model with correlated hopping on Bethe lattice. We calculated the one particle density of states (DOS) and two particle transport function (the ``quasiparticle'' scattering time) for a wide range of the correlated hopping parameter values and observed the singularities on the transport function due to the resonant two-particle contributions, whereas the one particle DOS does not show any anomalous features.
By tuning the doping of itinerant electrons, one can bring the chemical potential close to the resonant frequency and a large increase of the electrical and thermal conductivities and of the thermoelectric power can be achieved. At the same time, a strong enhancement of the Drude peak is developed on the optical conductivity spectra and a strong deviation from the Debye relaxation is observed at low temperatures. 
On the other hand, for some values of correlated hopping, when the hopping amplitude between the occupied sites is sufficiently reduced, itinerant electrons localize in the clusters of sites occupied by \(f\)-electrons giving rise to an additional narrow band in the DOS between the lower and upper Hubbard bands separated by an additional gap. 

It was already shown by Zlati\'c and Freericks~\cite{zlatic:266601} and Zlati\'c et al.~\cite{zlatic:155101} that light doping of the Mott insulator hugely enhanced its thermoelectric properties and electronic thermal transport, which displays a universal behaviour in the case of bad metals. However, nobody has investigated these for the systems with correlated hopping.
The main purpose of this article is to study the dc charge and thermal transport as well as thermopower for the Falicov-Kimball model with correlated hopping at the light doping levels.

The paper is organized as follows. 
In section~\ref{sec:formalism}, we present the DMFT solution for the Falicov-Kimball model with correlated hopping on a Bethe lattice and provide a derivation of the expressions for the charge and thermal transport coefficients.
In section~\ref{sec:results}, we present our results for the charge and thermal transport for different values of the correlated hopping parameters and for different doping levels. The results are summarized in section~\ref{sec:conclusions}. 

\section{Dynamical mean-field theory for thermoelectric transport on Bethe lattice with correlated hopping}\label{sec:formalism}

We consider the Falicov-Kimball model~\cite{falicov:997} with correlated hopping described by the Hamiltonian
\begin{equation}\label{eq:Hamiltonian}
H = H_{\textrm{loc}} + H_t\,,
\end{equation}
which contains two terms:
\begin{equation}\label{eq:Hloc}
H_{\textrm{loc}} = \sum_i \left[Un_{id}n_{if} - \mu_f n_{if} - \mu_d n_{id}\right]
\end{equation}
includes local correlations between the itinerant \(d\)-electrons and localized \(f\)-electrons and
\begin{equation}\label{eq:Ht}
H_t=\sum_{\langle ij\rangle} \frac{t_{ij}^{*}}{\sqrt{Z}} \Bigl[t_1d_i^{\dag}d_j +
t_2d_i^{\dag}d_j\left(n_{if}+n_{jf}\right) 
+ t_3d_i^{\dag}d_jn_{if}n_{jf}\Bigr]
\end{equation}
describes the nearest-neighbour inter-site hopping with amplitude \(t_1\) and nonlocal correlations, the so-called correlated hopping, with amplitudes \(t_2\) and \(t_3\) on the Bethe lattice with infinite coordination number, \(Z\rightarrow\infty\).
The occupation of localized \(f\) states is conserved, \([n_{if},H]=0\), and, by introducing the projection operators \(P_i^+=n_{if}\) and \(P_i^-=1-n_{if}\), one can define the projected \(d\)-electron operators
\begin{equation}
\bm{d}_i=\begin{pmatrix} d_i P_i^+ \\ d_i P_i^- \end{pmatrix}, 
\label{eq:proj_d}
\end{equation}
so that the nonlocal term can be rewritten in a compact matrix form~\cite{shvaika:075101}
\begin{align}
H_t&=\sum_{\langle ij\rangle} \frac{t_{ij}^{*}}{\sqrt{Z}} \Bigl[
t^{++}P_i^+d_i^{\dag}d_jP_j^+ + t^{--}P_i^-d_i^{\dag}d_jP_j^-
+t^{+-}P_i^+d_i^{\dag}d_jP_j^- +
t^{-+}P_i^-d_i^{\dag}d_jP_j^+\Bigr]
\nonumber\\
&= \sum_{\langle ij\rangle} \frac{t_{ij}^{*}}{\sqrt{Z}} \bm{d}_i^{\dag}\mathbf{t}\bm{d}_j.
\end{align}
Here, the hopping matrix 
\begin{align}
\mathbf{t}&=\begin{bmatrix}
t^{++} & t^{+-} \\
t^{-+} & t^{--}
\end{bmatrix} 
\label{eq:t_mtrx}
\end{align}
is defined in terms of the initial hopping amplitudes by
\begin{equation}
t^{--}=t_1\,, \qquad
t^{+-}=t^{-+}=t_1+t_2\,, \qquad
t^{++}=t_1+2t_2+t_3\,, 
\end{equation}
where \(t^{--}\), \(t^{+-}=t^{-+}\), and \(t^{++}\) describe hopping between the sites with different filling of \(f\)-states: both empty, one empty and one occupied, and both occupied, respectively.

Accordingly, the matrix Green's function for projected \(d\)-electrons \(\mathbf{G}_{ij}=[ {G}_{ij}^{\alpha\beta}]\), where \(\alpha,\beta=\pm\), is defined by 
\begin{equation}
\mathbf{G}_{ij}(\tau-\tau') = -\left\langle \mathcal{T} \bm{d}_i(\tau) \otimes \bm{d}_j^{\dagger}(\tau')\right\rangle \,, 
\end{equation}
where \(\mathcal{T}\) is the imaginary-time ordering operator and the angular bracket denotes the quantum statistical averaging with respect to \(H\). 
Due to the nonlocal character of correlated hopping, it is convenient to treat \(H_t\) as perturbation and expand around the atomic limit. The corresponding Dyson-type equation can be written in a matrix form as follows:
\begin{equation}\label{eq:Dyson}
\mathbf{G}_{ij}(\omega) = \bm{\Xi}_{ij}(\omega) + \sum_{\langle i'j'\rangle} \bm{\Xi}_{ij'}(\omega)  \cdot \frac{t_{j'i'}^{*}}{\sqrt{Z}} \mathbf{t} \cdot \mathbf{G}_{i'j}(\omega),
\end{equation}
where \(\bm{\Xi}_{ij}(\omega)\) is the irreducible cumulant~\cite{metzner:8549,georges:13}.

It can be shown that the irreducible cumulant is local in the limit of infinite coordination \(Z\to\infty\)~\cite{metzner:8549},
\(
\bm{\Xi}_{ij}(\omega) = \delta_{ij} \bm{\Xi}(\omega)
\), 
and can be calculated within the dynamical mean-field theory (DMFT). 
Now, the formal solution of the Dyson equation \eqref{eq:Dyson} for lattice Green's function can be written in a matrix form as follows:
\begin{equation}
\mathbf{G}_{\epsilon} (\omega)=\left[\bm{\Xi}^{-1}(\omega) - \mathbf{t}\epsilon\right]^{-1}	
\end{equation}
with the components
\begin{equation}
G_{\epsilon}^{\beta\alpha}(\omega) = \frac{A_{\beta\alpha}(\omega) - B_{\beta\alpha}\epsilon}{C(\omega) - D(\omega)\epsilon + \epsilon^2 \det\mathbf{t}} .
\label{GkvsE}
\end{equation}
Here, the band energy is distributed according to the density of states \(\rho(\epsilon)\), the semi-elliptic one for the Bethe lattice
\begin{equation}\label{eq:dos}
\rho(\epsilon) = \frac{2}{\piup W^2}\sqrt{W^2-\epsilon^2} \,, 
\end{equation}
and we introduced two adjugate matrices
\begin{equation}
\mathbf{A}(\omega) = \adj \bm{\Xi}^{-1}(\omega) = \bm{\Xi}(\omega)/\det\mathbf{\Xi}(\omega)
\end{equation}
and 
\begin{equation}
\mathbf{B} = \adj \mathbf{t} = \mathbf{t}^{-1}\det\mathbf{t} \,.
\end{equation}
In our case, the scalars \(C\) and \(D\) are given by 
\begin{equation}
C(\omega) = \det \mathbf{A}(\omega) = \det \bm{\Xi}^{-1}(\omega) = 1/\det \bm{\Xi}(\omega),
\end{equation} 
and
\begin{align}
D(\omega) = \Tr \left[\mathbf{A}(\omega)\mathbf{t}\right] = \Tr \left[\bm{\Xi}^{-1}(\omega)\mathbf{B}\right].
\end{align}

An irreducible cumulant \(\bm{\Xi}_{ij}(\omega)\) can be found as a solution of the DMFT equations
\begin{equation}   \label{eq:DMFT} 
\mathbf{G}_{\text{local}}(\omega)
\equiv\mathbf{G}_{ii}(\omega)=  \int\limits_{-\infty}^{+\infty}\rd\epsilon \rho(\epsilon)
\mathbf{G}_{\epsilon} (\omega) 
 = 
\left[\bm{\Xi}^{-1}(\omega) - \bm{\Lambda}(\omega)\right]^{-1} 
= \mathbf{G}_{\text{imp}}(\omega),
\end{equation}
where the local lattice Green's functions are equated with the one of an auxiliary impurity embedded in a self-consistent bath, described by the time-dependent mean field \(\bm{\Lambda}(\omega)=[\lambda^{\alpha\beta}(\omega)]\) (\(\lambda\)-field). For the Bethe lattice, we can rewrite the DMFT equation \eqref{eq:DMFT} as~\cite{shvaika:075101}
\begin{equation}\label{eq:DMFT_Bethe}
\bm{\Lambda}(\omega)=\frac{W^2}{4}\mathbf{t}\mathbf{G}_{\text{imp}}(\omega)\mathbf{t}
\end{equation}
and in numerical calculations we use \(W=2\), which defines our energy scale. 

For the Falicov-Kimball model with correlated hopping, the components of the Green's function of impurity are given by exact expressions~\cite{shvaika:075101}
\begin{align}
G_{\text{imp}}^{++}(\omega) & = w_1 g_1(\omega),
\nonumber \\
G_{\text{imp}}^{--}(\omega) & = w_0 g_0(\omega),
\nonumber\\
G_{\text{imp}}^{+-}(\omega) & = G_{\text{imp}}^{-+}(\omega) = 0,
\label{eq:Gimp}
\end{align}
where \(w_1=\langle P^+\rangle=\langle n_f\rangle\), \(w_0=\langle P^-\rangle=\langle 1- n_f\rangle\), and
\begin{align}
g_0(\omega) & =	\frac{1}{\omega+\mu_d - \lambda^{--}(\omega)}\,,
\nonumber \\
g_1(\omega) & = \frac{1}{\omega+\mu_d - U - \lambda^{++}(\omega)}
\label{eq:gimp0}
\end{align}
are the impurity Green's functions of a conduction electron in the presence of a \(f\)-state which is either permanently empty or occupied, respectively. After substitution of these expressions in \eqref{eq:DMFT_Bethe}, one can get the 4th order polynomial equations for \(g_0(\omega)\) or \(g_1(\omega)\), and details of its solution are given in \cite{dobushovskyi:125133}.
 
For the local single-particle Green's function 
\begin{equation}
G_{ii}(\tau-\tau') = -\left\langle \mathcal{T} d_i(\tau) d_i^{\dagger}(\tau')\right\rangle ,
\end{equation}
we have 
\begin{equation}
G_{ii}(\omega) = \sum_{\alpha,\beta=\pm} G_{\text{imp}}^{\alpha\beta}(\omega) = w_0 g_{0}(\omega) + w_1 g_{1}(\omega) ,
\label{eq:G_ii}
\end{equation}
and, finally, the renormalized DOS of the lattice is expressed in terms of the impurity Green's function
\begin{equation}
A_d(\omega) = -\frac{1}{\piup} \Img G_{ii}(\omega)
= -\frac{1}{\piup} \left[w_0 \Img g_{0}(\omega) + w_1 \Img g_{1}(\omega)\right].
\end{equation}

The chemical potential for \(d\)-electrons \(\mu_d\) is obtained by solving the equation
\begin{equation}
n_d = -\frac{1}{\piup} \int_{-\infty}^{+\infty} \rd\omega f(\omega) \Img G_{ii}(\omega),
\end{equation}
where \(f(\omega)=1/(\re^{\omega/T}+1)\) is the Fermi function, for a given value of their concentration \(n_d=\langle n_d\rangle\).


We now proceed to the calculation of transport properties by linear response theory. 
The dc charge conductivity 
\begin{equation}\label{eq:dc}
\bm{\sigma}_{\textrm{dc}} = e^2 \mathbf{L}_{11}\, ,
\end{equation}
the Seebeck coefficient (thermoelectric power \(\bm{E}=\mathbf{S}\nabla T\)) 
\begin{equation}\label{thermcond}
\mathbf{S} =  \frac{1}{eT} \mathbf{L}_{11}^{-1} \mathbf{L}_{12}\,,
\end{equation}
and the electronic contribution to thermal conductivity 
\begin{equation}\label{eq:thermcond}
\bm{\kappa}_{\textrm{e}} =  \frac1T \left[\mathbf{L}_{22} - \mathbf{L}_{21} \mathbf{L}_{11}^{-1} \mathbf{L}_{12}\right]
\end{equation}
are expressed in terms of the transport integrals \cite{jonson:4223,jonson:9350,freericks:035133,shvaika:43704}
\begin{equation}\label{JM}
\mathbf{L}_{lm} = \frac{\sigma_0}{e^2} \int_{-\infty}^{+\infty} \rd \omega \left[-\frac{\rd f(\omega)}{\rd \omega}\right] \mathbf{I}(\omega) \omega^{l+m-2},
\end{equation}
where \(\mathbf{I}(\omega)\) is the transport function.  
In the considered case of correlated hopping, the DMFT expression for transport function reads
\begin{align}
I(\omega) &= \frac{1}{\piup} \int \rd \epsilon \rho(\epsilon) \Phi_{xx}(\epsilon) \Tr \left[\mathbf{t}\, \Img \mathbf{G}_{\epsilon} (\omega)\,  \mathbf{t}\, \Img \mathbf{G}_{\epsilon}(\omega)\right]
\nonumber\\
&= \frac{1}{\piup} \sum_{\alpha\beta\alpha'\beta'} t^{\alpha\beta} t^{\alpha'\beta'} \int \rd \epsilon \rho(\epsilon) \Phi_{xx}(\epsilon) \Img G_{\epsilon}^{\,\beta\alpha'} (\omega) \Img G_{\epsilon}^{\,\beta'\alpha}(\omega), 
\label{eq:transport}
\end{align}
where \(\Phi_{xx}(\epsilon)\)  is the so-called  lattice-specific transport DOS~\cite{arsenault:205109} and, for the \(Z=\infty\) Bethe lattice with semielliptic DOS,  
the \(f\)-sum rule yields~\cite{chung:11955}
\begin{equation}
\Phi_{xx}(\epsilon) = \frac{1}{3Z}\left(W^2-\epsilon^2\right) .
\label{eq:Phi_Bethe}
\end{equation}

Finally, the transport function reads~\cite{dobushovskyi:125133}
\begin{align}
I(\omega)&=\frac{1}{2\piup} \Biggl( \Real \left\{\Psi^{\prime}[E_1(\omega)]+\Psi^{\prime}[E_2(\omega)] \right\} - \frac{\Img \Psi\left[E_1(\omega)\right] }{\Img E_1(\omega)}-\frac{\Img \Psi\left[E_2(\omega)\right] }{\Img E_2(\omega)}
\nonumber\\
&- K(\omega)\Biggl\lbrace\frac{1}{\Img E_1(\omega)}\Img\frac{ \Psi[E_1(\omega)] } {\left[E_1(\omega)-E_2(\omega)\right]\left[E_1(\omega)-E^*_2(\omega)\right]} 
\nonumber\\
&+\frac{1}{\Img E_2(\omega)}\Img \frac{ \Psi[E_2(\omega)] } {\left[E_2(\omega)-E_1(\omega)\right]\left[E_2(\omega)-E^*_1(\omega)\right]}\Biggr\rbrace
\Biggr),
\label{eq:Iw}
\end{align}
where \(E_{1}\) and \(E_{2}\) are the roots of the denominator in equation~\eqref{GkvsE}, 
\(C(\omega) - D(\omega)\epsilon + \epsilon^2 \det\mathbf{t}=0\), given by  
\begin{align}
E_{1}(\omega) &= \frac{D(\omega)}{2\det \mathbf{t}} \left[ 1 + \sqrt{1 - \frac{4C(\omega)}{D^2(\omega)}\det \mathbf{t}}\, \right],
\label{eq:E1}
\\
E_{2}(\omega) &= \frac{2C(\omega)}{D(\omega)} \left[ 1 + \sqrt{1 - \frac{4C(\omega)}{D^2(\omega)}\det \mathbf{t}}\, \right]^{-1},
\label{eq:E2}
\end{align}
and \(K(\omega)\) reads
\begin{equation}
K(\omega)=2 \Real [E_1(\omega)E^*_2(\omega)]-\frac{1}{\det \mathbf{t}}\Real \Tr [\mathbf{A}^*(\omega) \bm{\Xi}^{-1}(\omega)].
\end{equation}
Here,
\begin{align}
\Psi(\zeta)&=\int \rd \epsilon \frac{\rho(\epsilon)}{\zeta - \epsilon} \Phi_{xx}(\epsilon) ,
\nonumber \\
\Psi^{\prime}(\zeta)&=\frac{\rd \Psi(\zeta)}{\rd \zeta}\,,
\end{align}
and, for the semielliptic DOS, we find
\begin{align}
\Psi(\zeta)&=\frac{1}{3}\left[(W^2 -\zeta^2) F(\zeta)+ \zeta \right] ,
\nonumber \\
\Psi^{\prime}(\zeta)&=\frac{1}{3}\left[(W^2 -\zeta^2) F^{\prime}(\zeta)+1-2 \zeta F(\zeta) \right] ,
\end{align}
where 
\begin{align}
F(\zeta) &= \int \rd \epsilon \frac{\rho(\epsilon)}{\zeta - \epsilon}= \frac{2}{W^2}\left(\zeta - \sqrt{\zeta^2 -W^2}\right) ,
\nonumber\\
F^{\prime}(\zeta) &=  \frac{\rd F(\zeta)}{\rd \zeta}=\frac{\zeta F(\zeta)-2 }{\zeta^2 - W^2}.
\end{align}

In addition, we calculate the Lorenz number
\begin{equation}\label{eq:LN}
L=\frac{\kappa_{\textrm{e}}}{\sigma_{\textrm{dc}}T}=\frac{1}{e^2T^2}\left[\frac{L_{22}}{L_{11}}-\left(\frac{L_{12}}{L_{11}}\right)^2\right],
\end{equation}
which for the pure metal with degenerate fermions is equal to \(L_0=\piup^2/3\), as it follows from the Wiedemann-Franz law.

\section{Results and discussion}\label{sec:results}

According to equations \eqref{eq:dc}--\eqref{JM}, the transport coefficients are determined by the shape and value of the transport function \(I(\omega)\) within the so-called Fermi window, defined by function \([-\rd f(\omega)/\rd\omega]\), around the Fermi level (chemical potential value)~\cite{shvaika:43704}. The largest thermoelectric effect can be observed when transport function is strongly asymmetric at the chemical potential, e.g., it is nonzero on the one side from the chemical potential and zero on the other side. Such case can be achieved in a lightly doped Mott insulators when the chemical potential is stuck at the band edge~\cite{zlatic:266601}.

\begin{figure}[!t]
	\centering
	\includegraphics[width=0.47\linewidth]{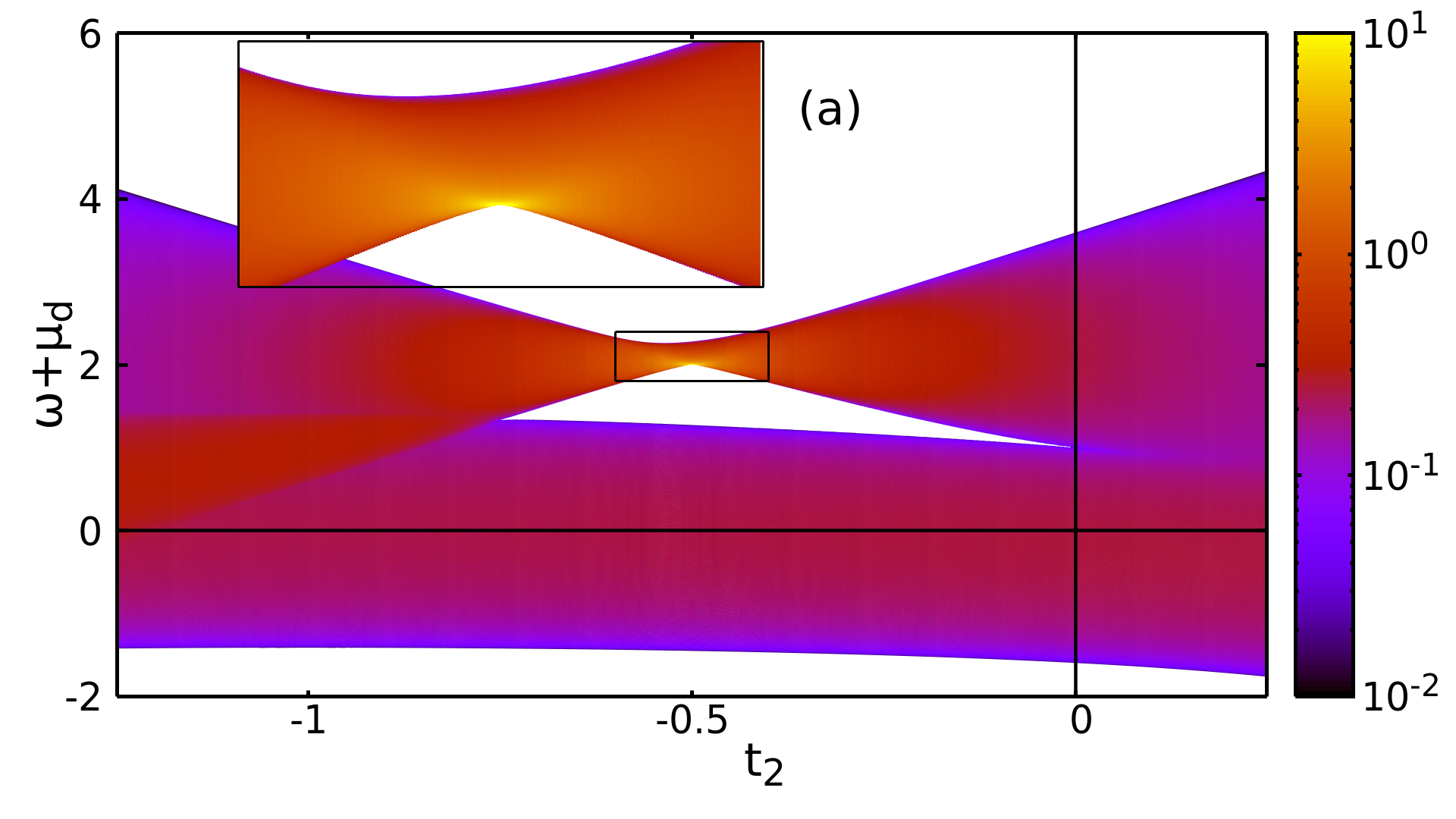}\qquad
	\includegraphics[width=0.47\linewidth]{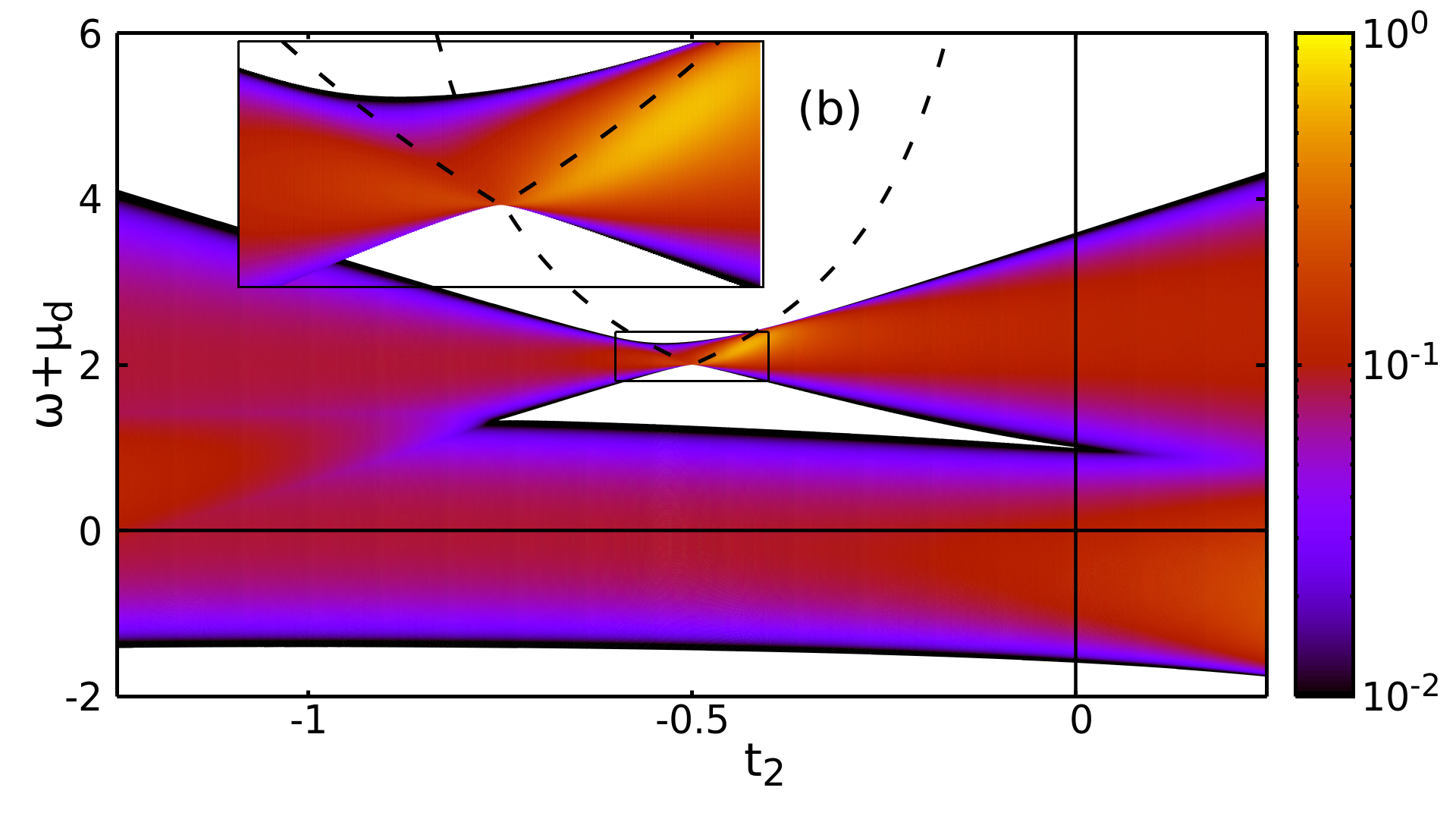} \\
	\includegraphics[width=0.47\linewidth]{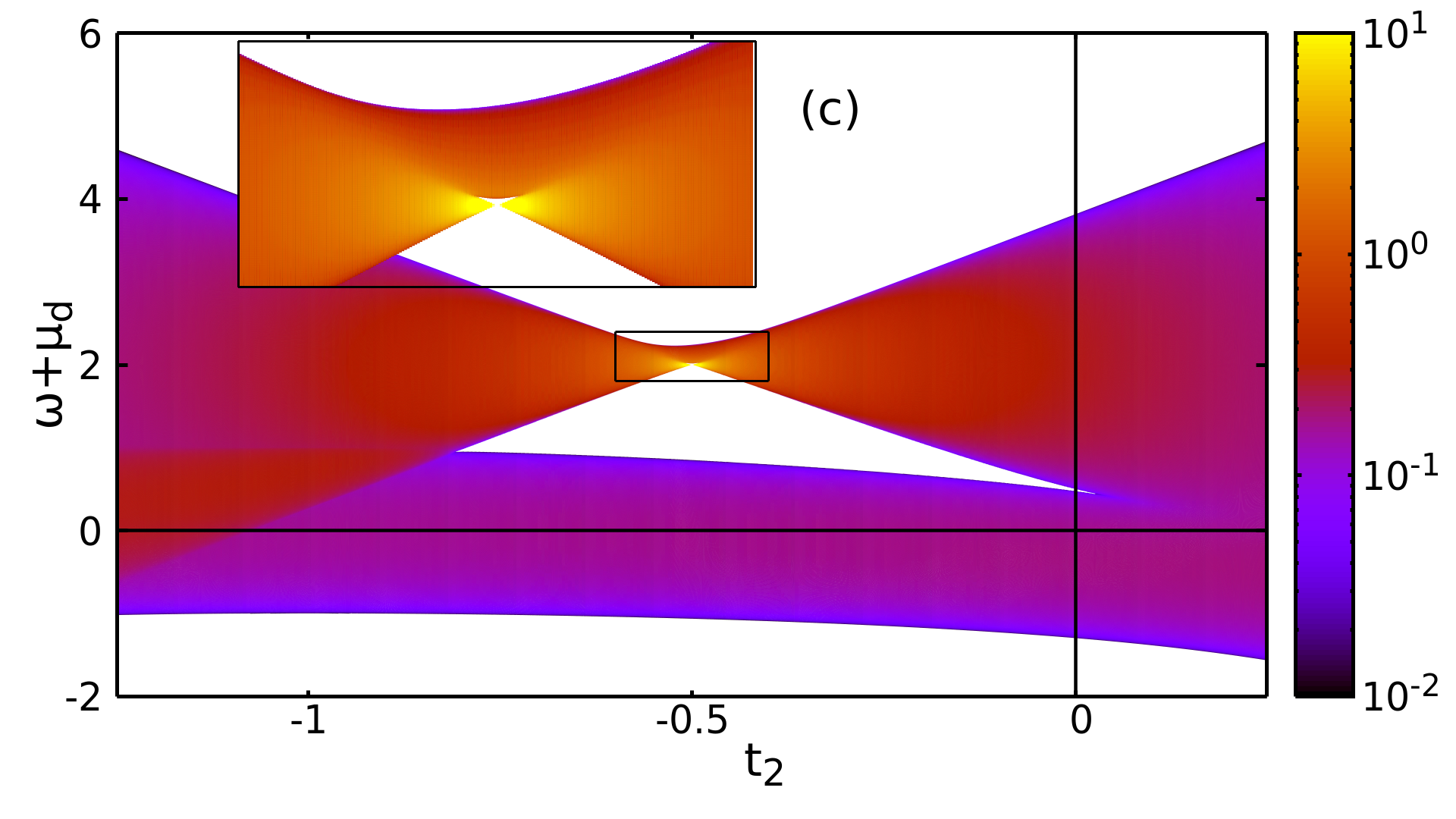}\qquad
	\includegraphics[width=0.47\linewidth]{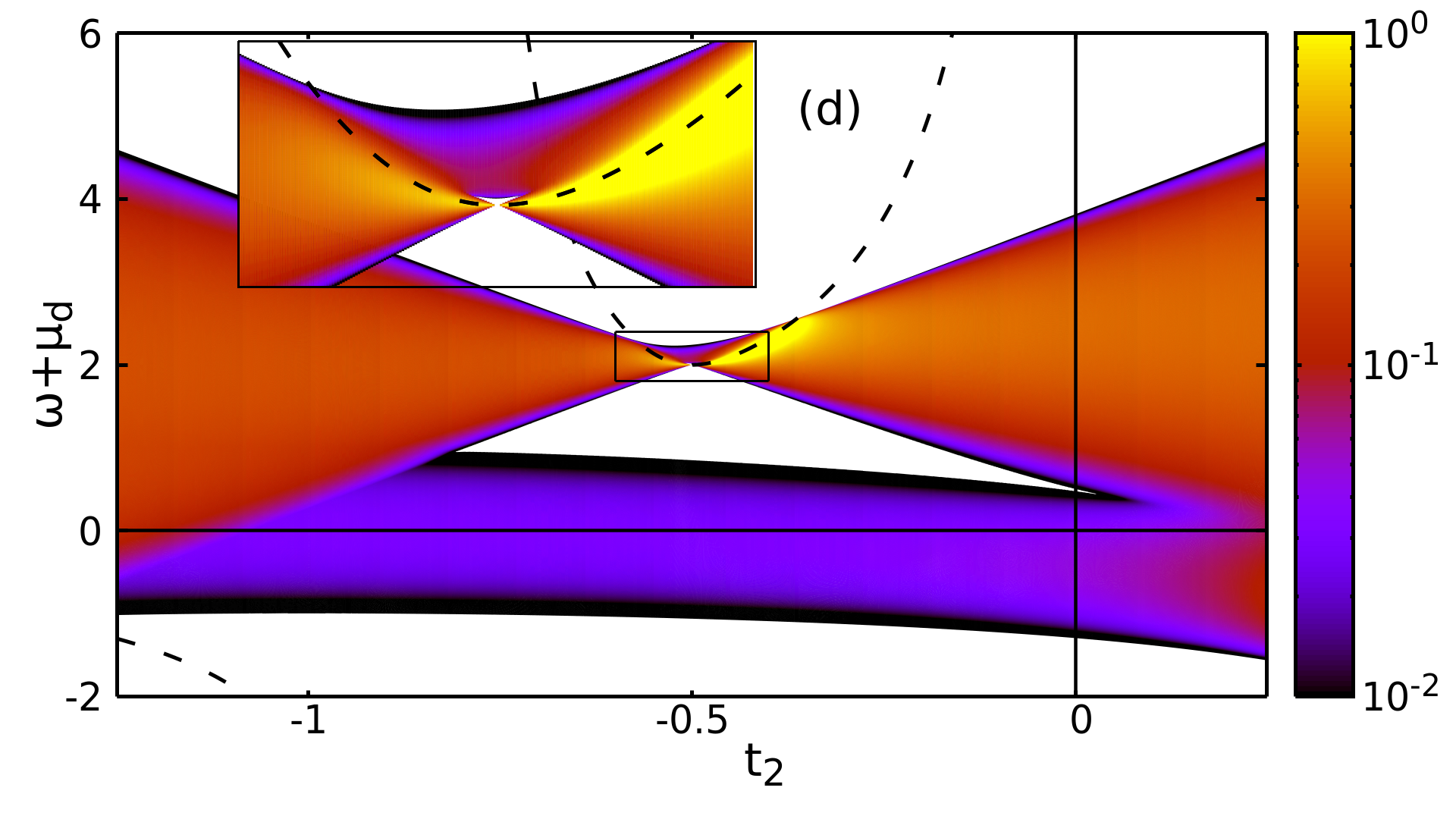} 
	\caption{(Colour online) (a, c) Density of states \(A_d(\omega)\) and (b, d) transport function \(I(\omega)\) for \(U=2.0\) and (a, b) \(n_f=0.5\) and (c, d) \(n_f=0.75\) depending on the correlated hopping parameter \(t_2\).}
	\label{fig:nf05}
	\label{fig:nf075}
\end{figure}

\begin{figure}[!t]
	\centering
	\includegraphics[width=0.45\linewidth]{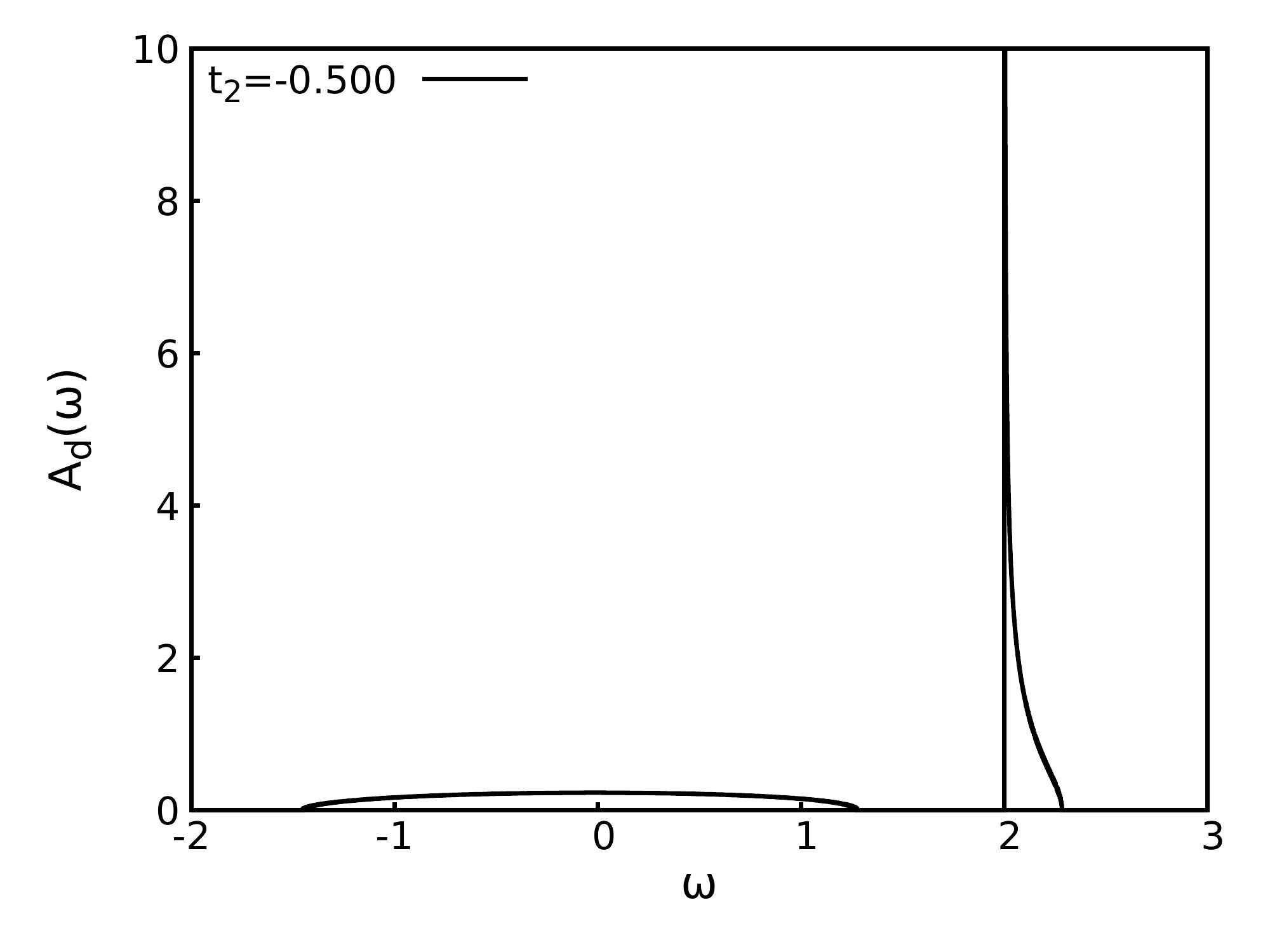}\qquad
	\includegraphics[width=0.45\linewidth]{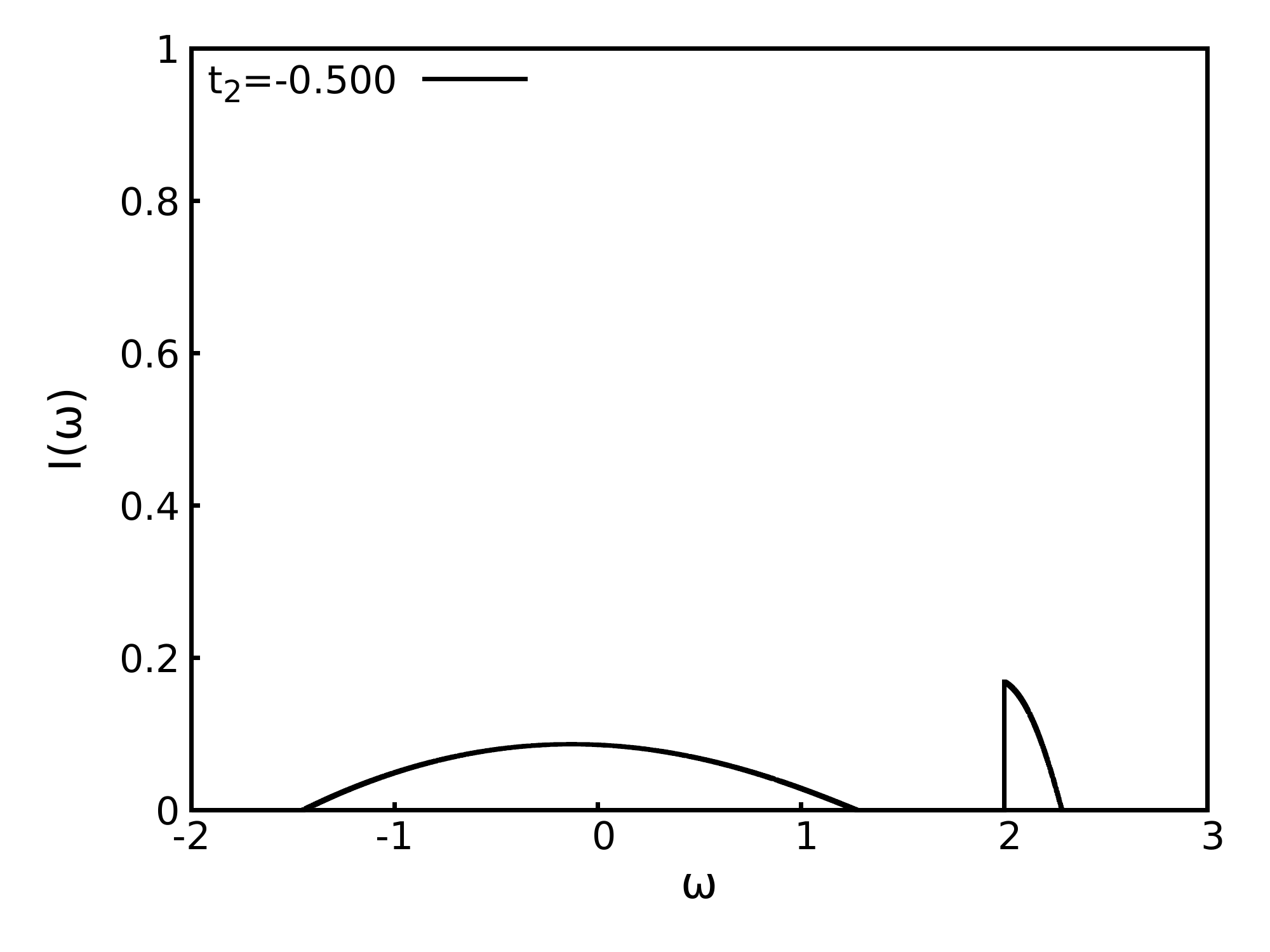} 
	\caption{Density of states \(A_d(\omega)\) and transport function \(I(\omega)\) for \(U=2.0\), \(n_f=0.5\), and \(t_2=-0.5\) (\(t^{++}=0\)).}
	\label{fig:dos-nf05t2-05}
\end{figure}

In the case of correlated hopping, the Mott insulator phase has some specific features. First of all, the DOS \(A_d(\omega)\) is strongly asymmetric and contains two Hubbard bands, the lower and the upper one, with spectral weights \(w_0=1-n_f\) and \(w_1=n_f\), respectively. The Mott gap is the largest at \(t_2=-0.5\) value, see figure~\ref{fig:nf05}~(a), when \(t^{++}=0\) and hopping between the sites with occupied \(f\) states is prohibited (here and below we put \(t_1=1\) and consider the case of \(t_3=0\)). Besides, at half-filling, \(n_f=0.5\), and exactly at \(t_2=-0.5\) value, there is a square root singularity at the lower edge of the upper Hubbard band, figure~\ref{fig:dos-nf05t2-05}, which strongly effects the temperature behaviour of the chemical potential. At high temperatures, the chemical potential \(\mu_d\) is placed closer to the lower Hubbard band, but with the temperature decrease it starts to approach the centre of the Mott gap.

\begin{figure}[!t]
	\centering
	\includegraphics[width=0.44\linewidth]{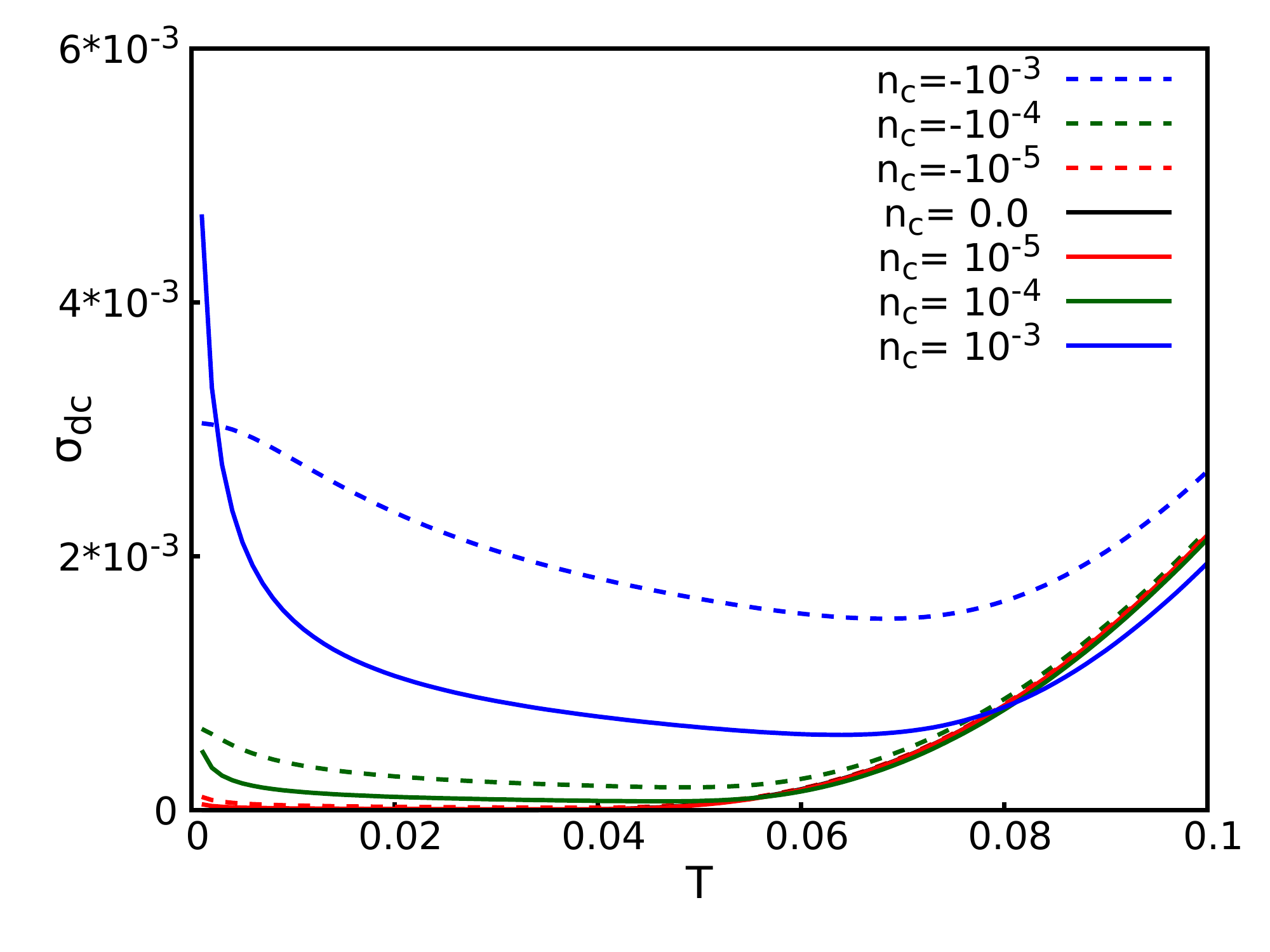} \qquad
	\includegraphics[width=0.44\linewidth]{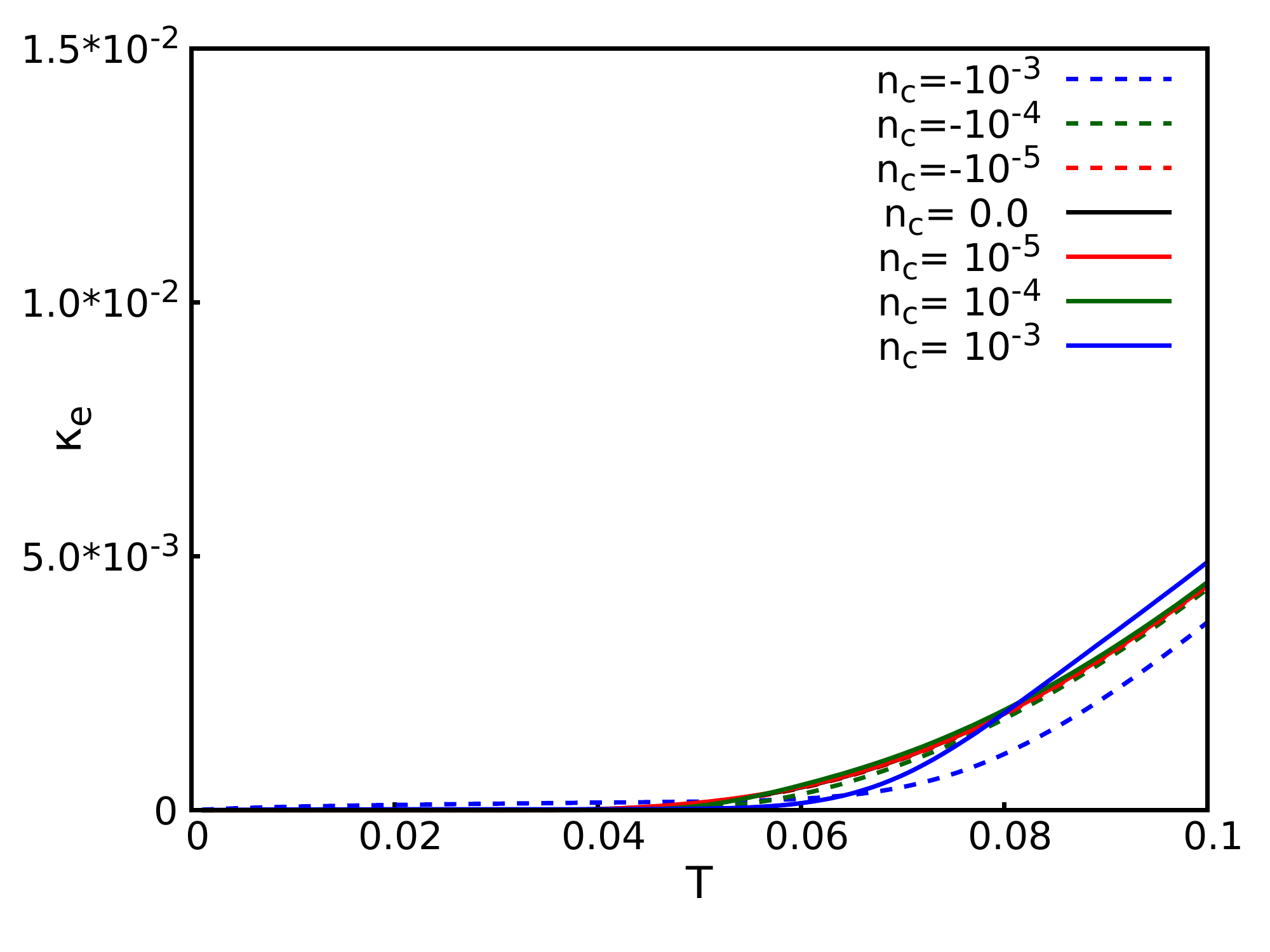} \\
	\includegraphics[width=0.44\linewidth]{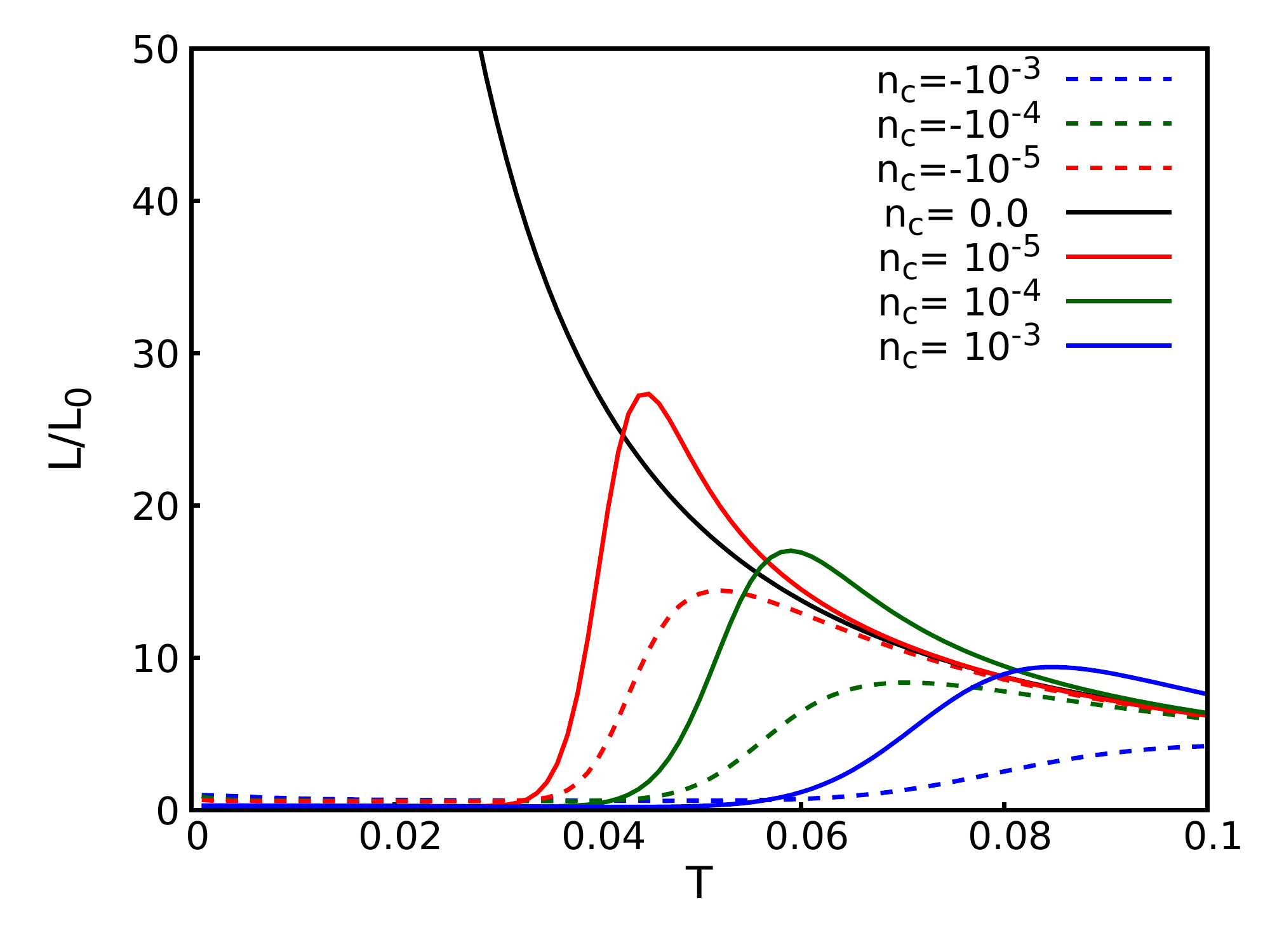} \qquad
	\includegraphics[width=0.44\linewidth]{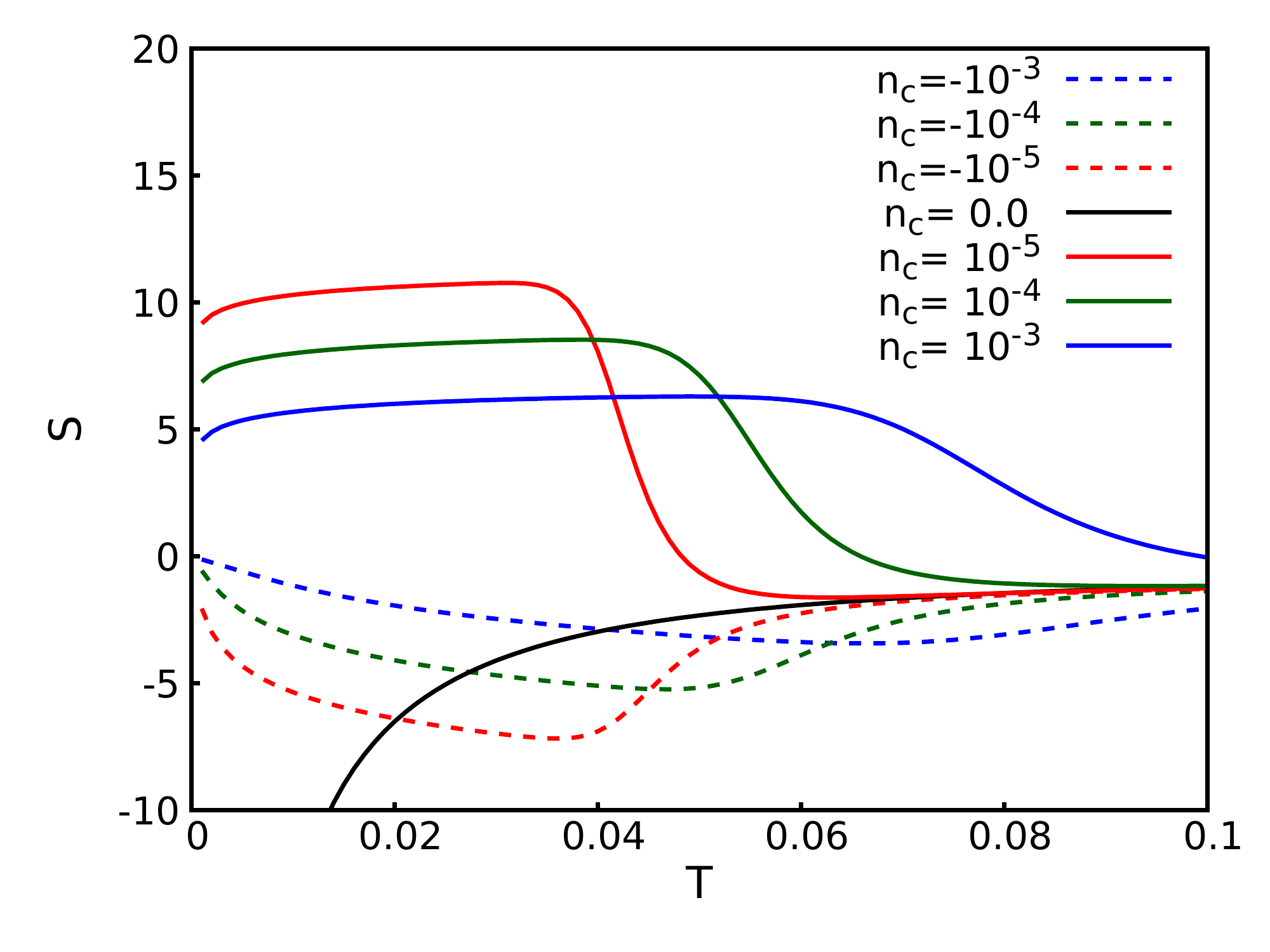} 
	\caption{(Colour online) Temperature dependences of the dc charge \(\sigma_{\textrm{dc}}\) and thermal \(\kappa_{\textrm{e}}\) conductivities, Lorenz number \(L\) (\(L_0=\piup^2/3\)), and thermopower (Seebeck coefficient) \(S\) for \(U=2.0\), \(n_f=0.5\), \(t_2=-0.5\) (\(t^{++}=0\)), \(n_d=0.5+n_c\).}
	\label{fig:therm-nf05t2-05}
\end{figure}

The shape of transport function \(I(\omega)\) is different and it is strongly affected by the resonant peak~\cite{dobushovskyi:125133} placed at the frequency \(\omega=\omega_{\text{res}}\) given by
\begin{equation}
\omega_{\text{res}} + \mu_d = \frac{U}{1-\eta} 
\label{eq:w_res}
\end{equation}
with
\begin{equation}\label{eq:x_res}
\eta = \frac{(t^{+-})^2}{(t^{--})^2} 
- \frac{(t^{+-})^2-\sqrt{(t^{+-})^4+4w_1w_0 \left[(t^{++}t^{--})^2-(t^{+-})^4\right] }}{2(t^{--})^2w_0}.
\end{equation}
Transport function displays a power law frequency dependence at the top of the lower Hubbard band~\cite{yamamoto:155201}, instead of the square root one for DOS, and an anomalous step-like feature at the bottom of the upper Hubbard band. As a result, the temperature dependences of transport coefficients, figure~\ref{fig:therm-nf05t2-05}, strongly depend on the doping character and on its level. At half filling \(n_d=0.5\) and in the Mott insulator phase, the dc charge and thermal conductivities display typical dependences for the large gap insulators with an exponential decay at \(\beta=1/T\to+\infty\) with divergent Lorenz number \(L\). On the other hand, the Seebeck coefficient is negative and follows the \(1/T\) dependence at small temperatures, which is caused by the above mentioned strong asymmetry of the transport function, power law dependence for the lower Hubbard band and step-like feature for the upper one, see, for comparison,~\cite{joura:165105}.

Now, let us consider the effect of doping. For the hole doping, \(n_d=0.5+n_c\) and \(n_c<0\), the chemical potential is placed at \(T=0\) somewhere at the top of the lower Hubbard band. The temperature dependences of the dc charge and thermal conductivities at high temperatures display the same behaviour as in the Mott insulator case. For low temperatures, when the chemical potential enters the lower Hubbard band, the dc charge conductivity starts to increase, as it should be for bad metal, and the thermal one displays the linear temperature dependence. The strongest doping effect is observed for the thermoelectric transport. Since now the chemical potential approaches the lower Hubbard band with the temperature decreasing faster than in the Mott insulator case, the Seebeck coefficient also increases faster until the chemical potential enters the lower Hubbard band. Then, the transport function becomes smooth within the Fermi window resulting in the lowering of the absolute value of Seebeck coefficient at low temperatures. We can notice a strong deviation of the temperature dependence of thermopower \(S(T)\) from the one typical of bad metals~\cite{zlatic:155101} and specific for light doping~\cite{zlatic:266601}. Now, \(S(T)\) exhibits almost a linear temperature dependence below the peak, which is caused by the character of the temperature dependence of the chemical potential due to the sharp features of DOS.  Similar behaviour is observed for an electron doping, \(n_c>0\), but now the Seebeck coefficient becomes positive when the chemical potential approaches the upper Hubbard band, and transport coefficients are larger in comparison with the hole doping case because the transport function is larger for the upper Hubbard band in comparison with the lower one. Moreover, the linear segment on \(S(T)\) extends and becomes flattened.

\begin{figure}[!t]
	\centering
	\includegraphics[width=0.45\linewidth]{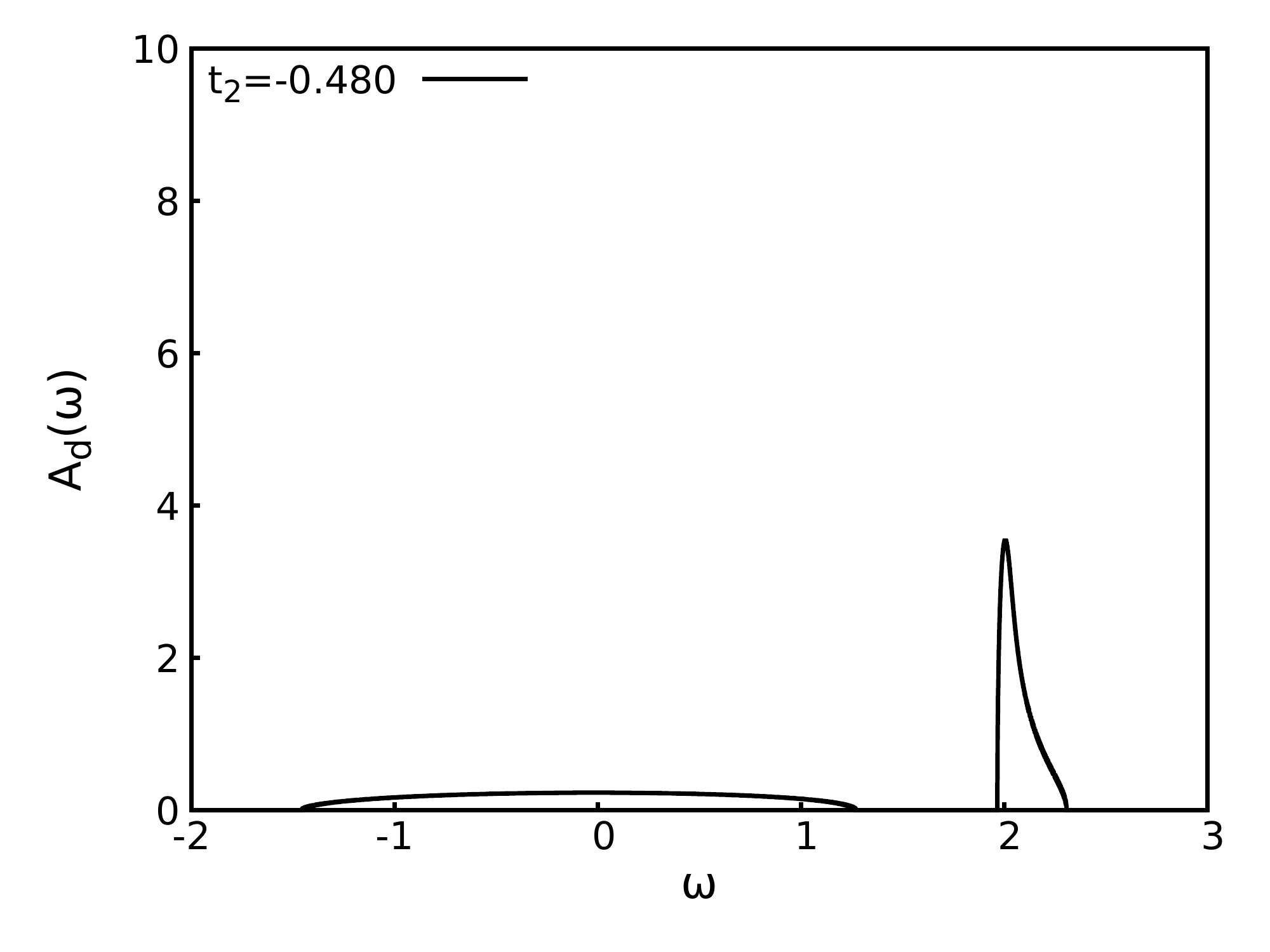}\qquad
	\includegraphics[width=0.45\linewidth]{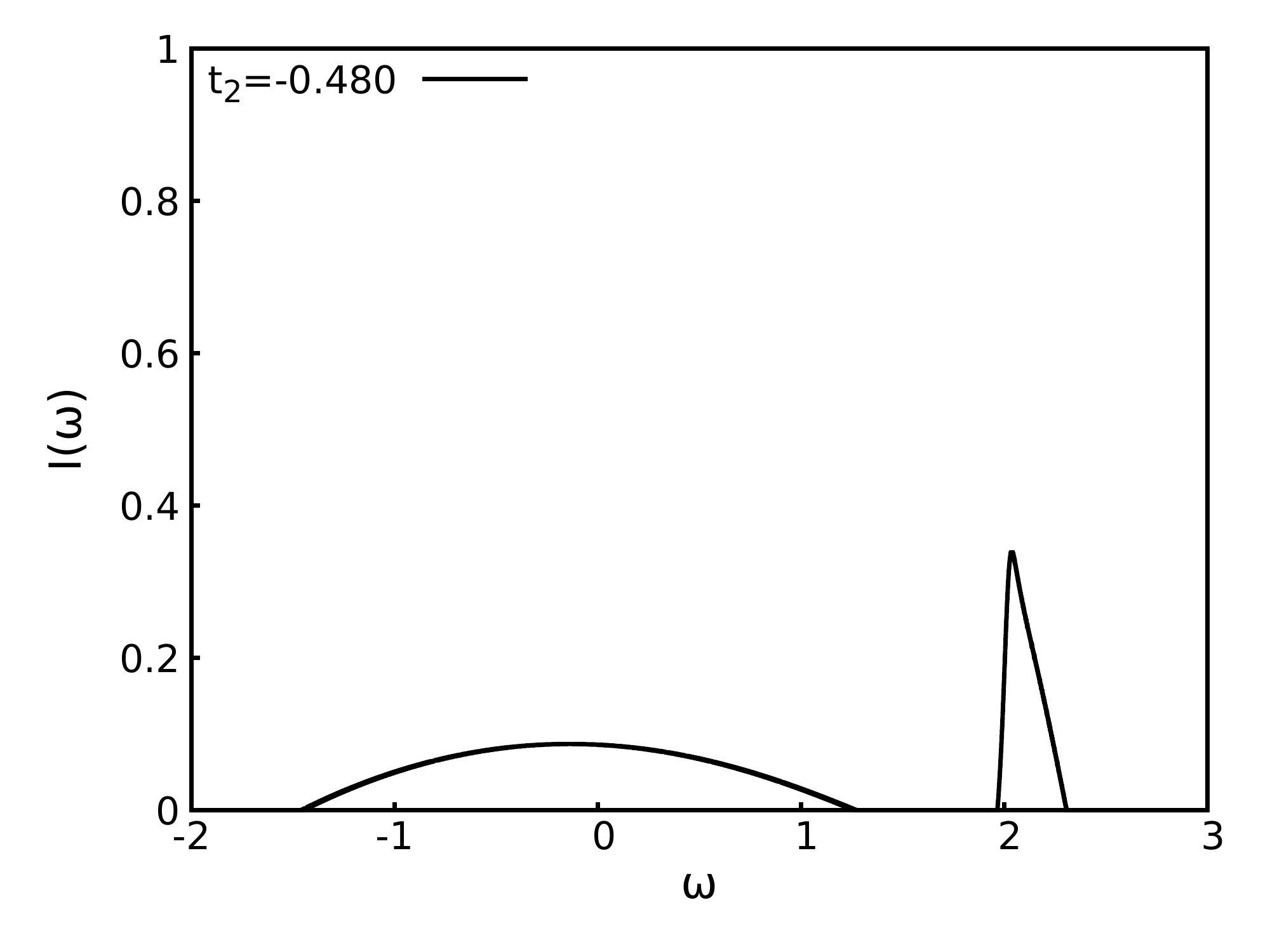} 
	\caption{Density of states \(A_d(\omega)\) and transport function \(I(\omega)\) for \(U=2.0\), \(n_f=0.5\), and \(t_2=-0.48\) (\(t^{++}=0.04\)).}
	\label{fig:dos-nf05t2-048}
\end{figure}


\begin{figure}[!t]
	\centering
	\includegraphics[width=0.45\linewidth]{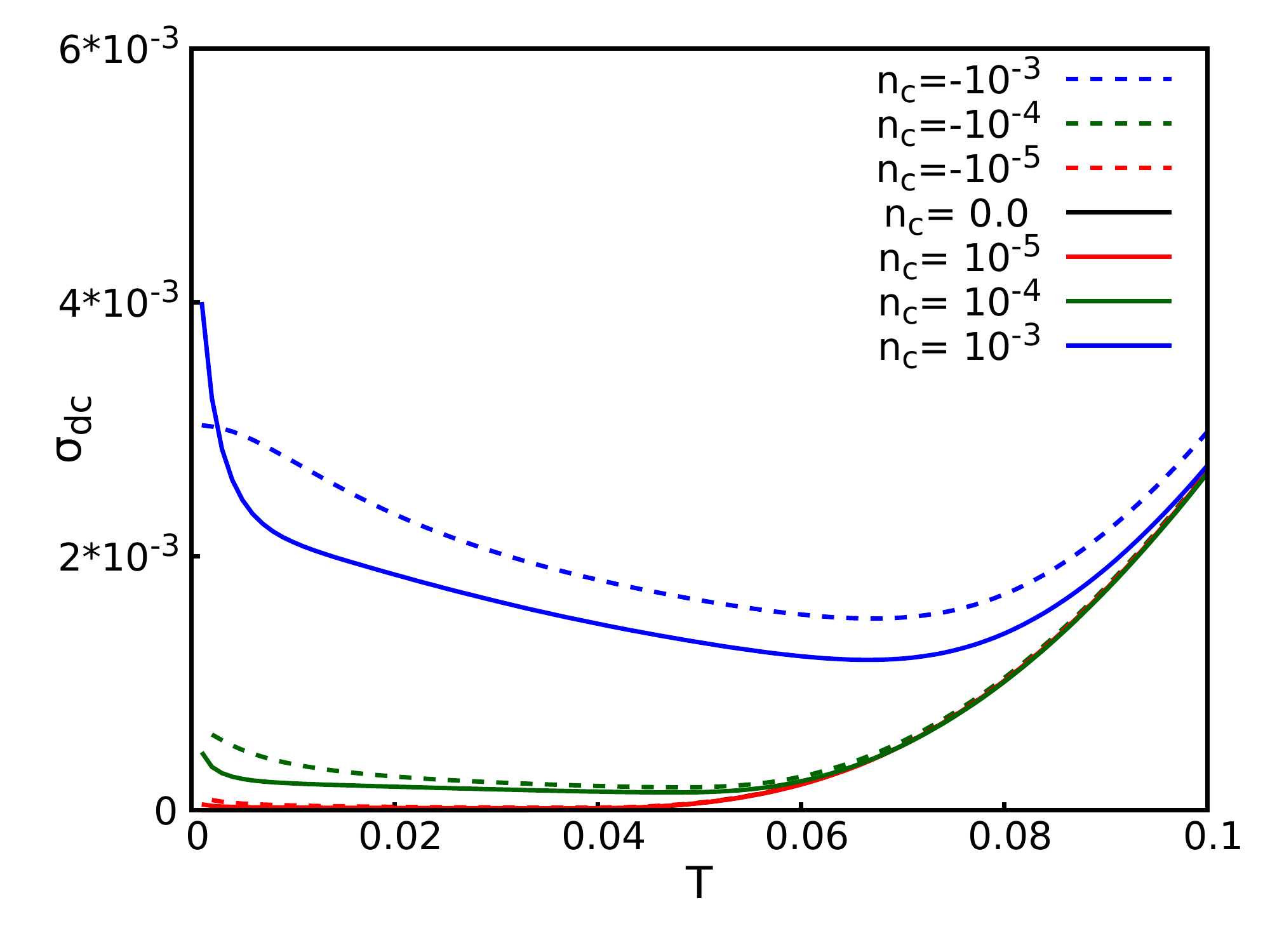} \qquad
	\includegraphics[width=0.45\linewidth]{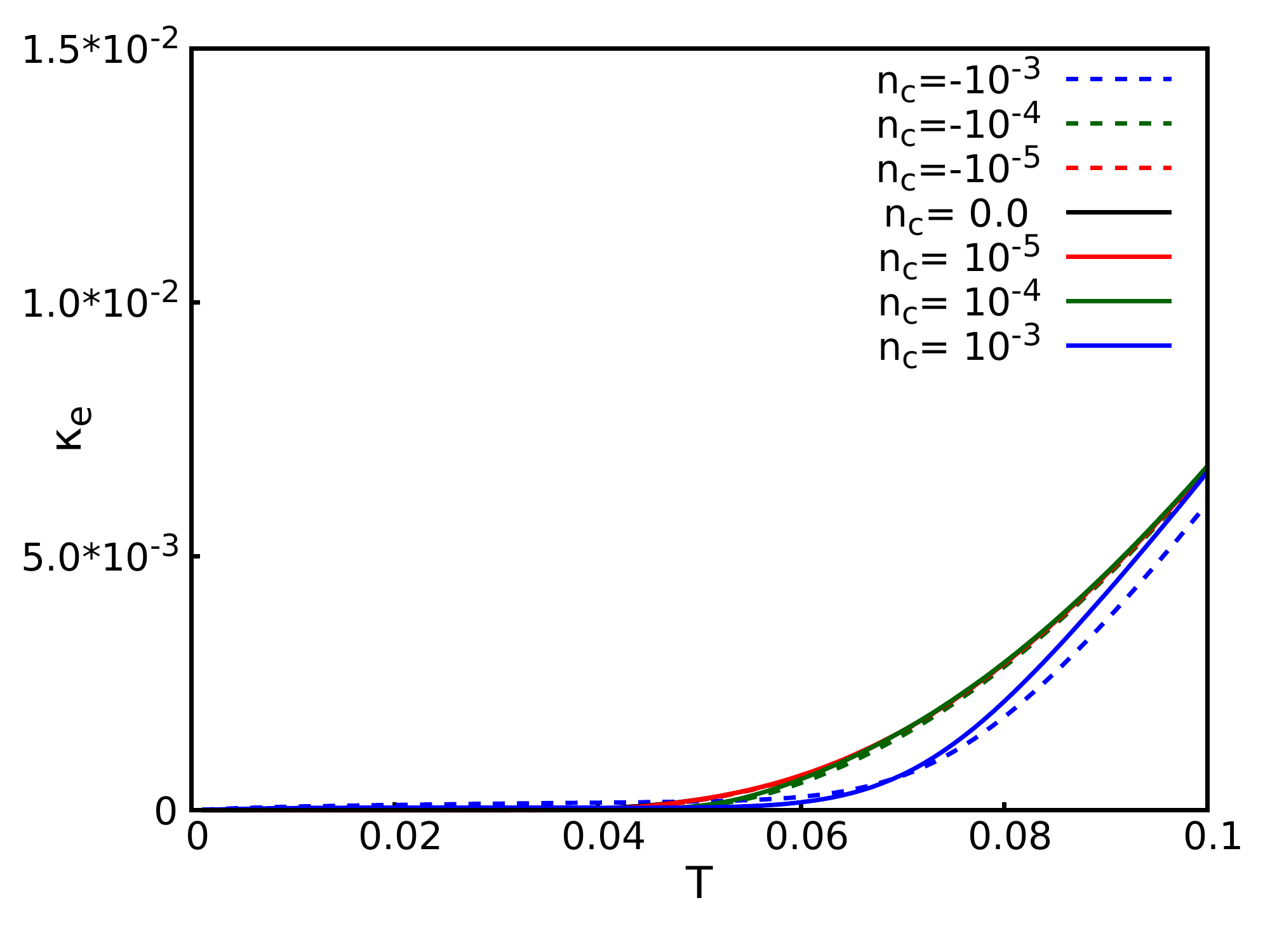} \\
	\includegraphics[width=0.45\linewidth]{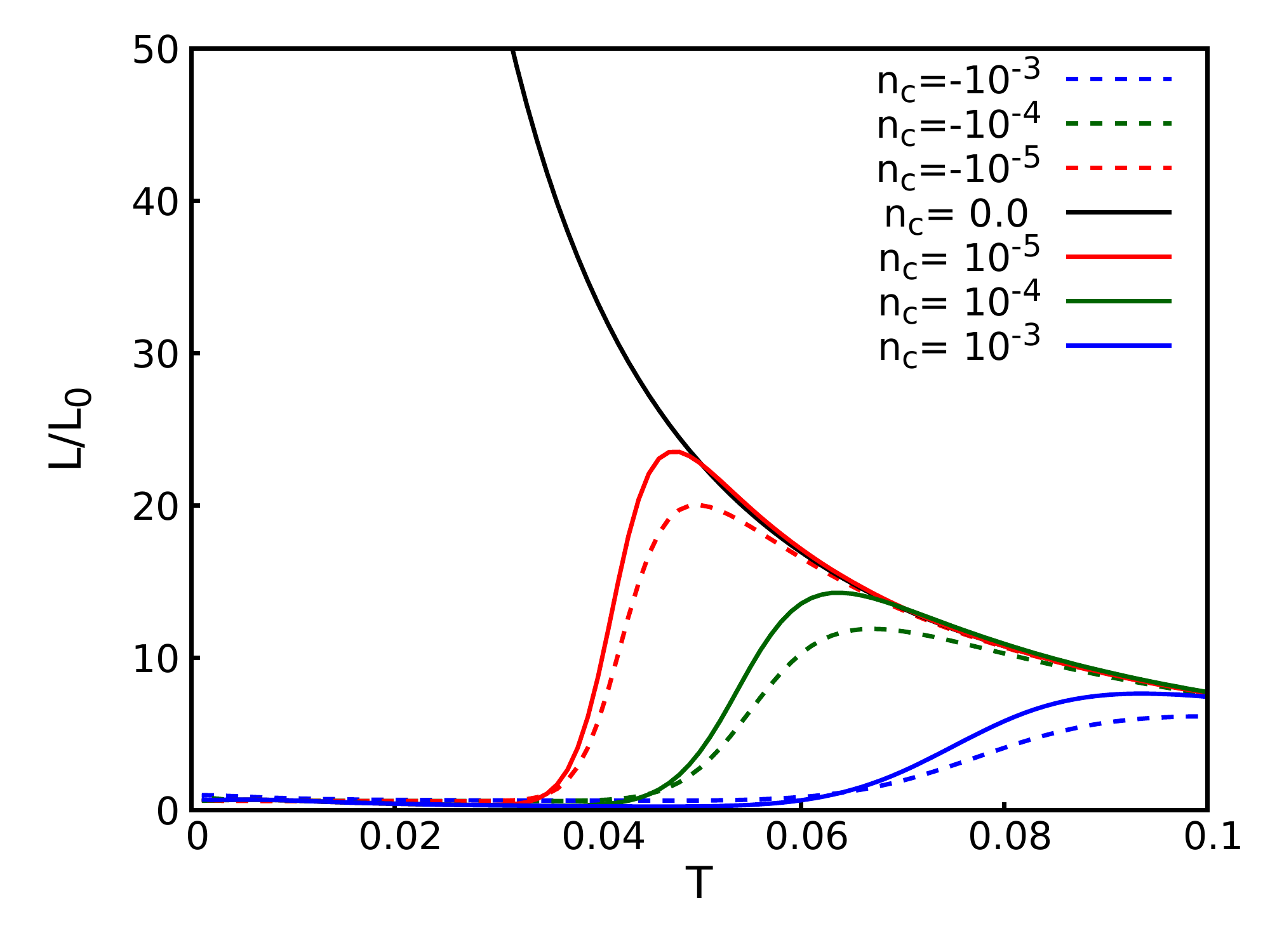} \qquad
	\includegraphics[width=0.45\linewidth]{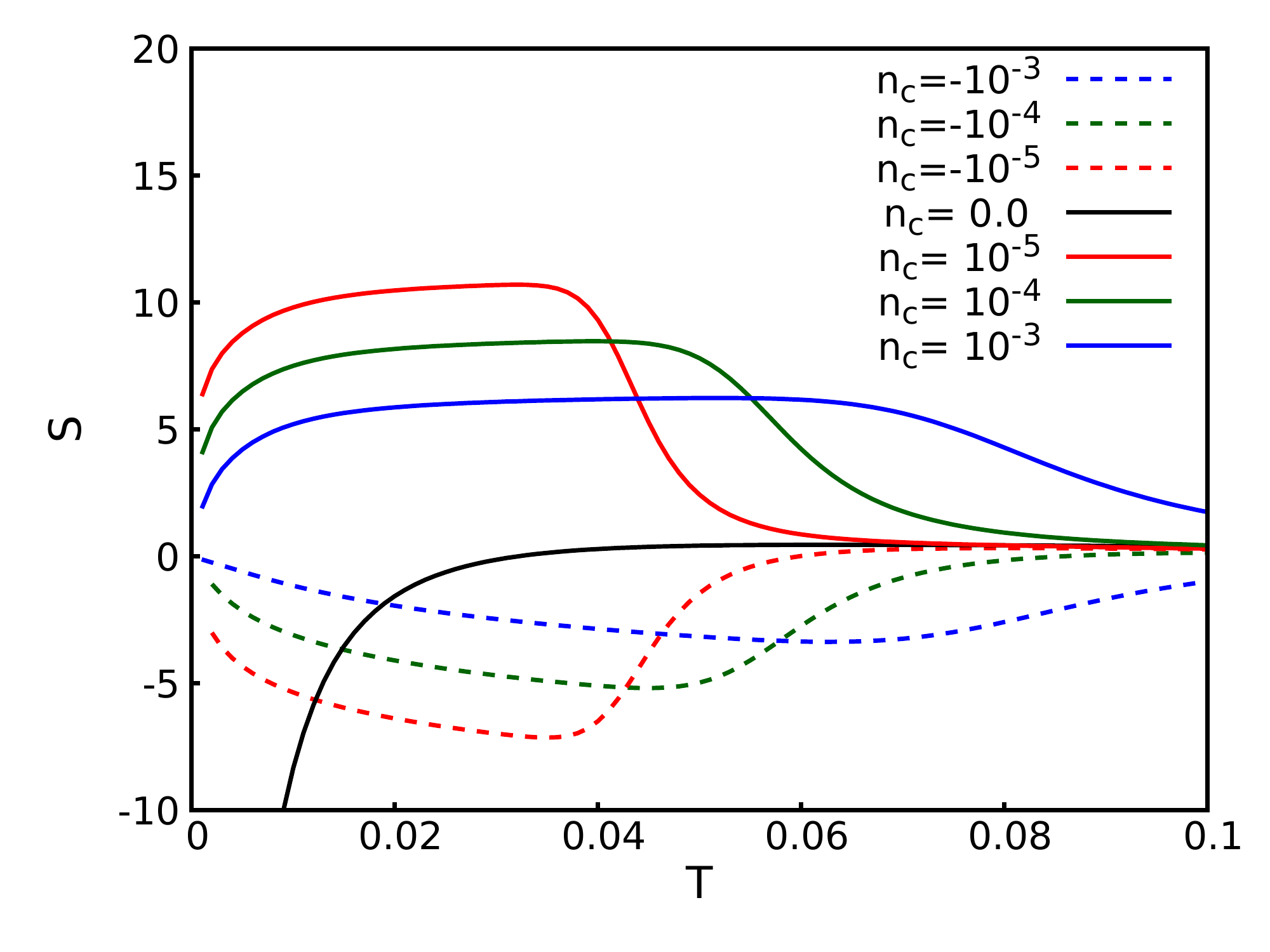} 
	\caption{(Colour online) Temperature dependences of the dc charge \(\sigma_{\textrm{dc}}\) and thermal \(\kappa_{\textrm{e}}\) conductivities, Lorenz number \(L\) (\(L_0=\piup^2/3\)), and thermopower (Seebeck coefficient) \(S\) for \(U=2.0\), \(n_f=0.5\), \(t_2=-0.48\) (\(t^{++}=0.04\)), \(n_d=0.5+n_c\).}
	\label{fig:therm-nf05t2-048}
\end{figure}

The results presented above were obtained for the special case of \(t_2=-0.5\), when the hopping between the sites with occupied \(f\) states is zero, \(t^{++}=0\). In figures~\ref{fig:dos-nf05t2-048}--\ref{fig:therm-nf05t2-048}, we present the results for the case with small values of \(t^{++}=0.04\). Now, the edge singularity on DOS is smoothed in a narrow peak, whereas the step-like feature on the transport function is replaced by the resonant peak. Nevertheless, the above discussed features are preserved, and we observe only quantitative changes in the temperature dependences of the transport coefficient. The only prominent effect is visible for the electron doping case, \(n_c>0\), when the step-like feature on the transport function is smoothed which leads to the faster decreasing of the Seebeck coefficient at \(T\to0\).

\begin{figure}[!t]
	\centering
	\includegraphics[width=0.45\linewidth]{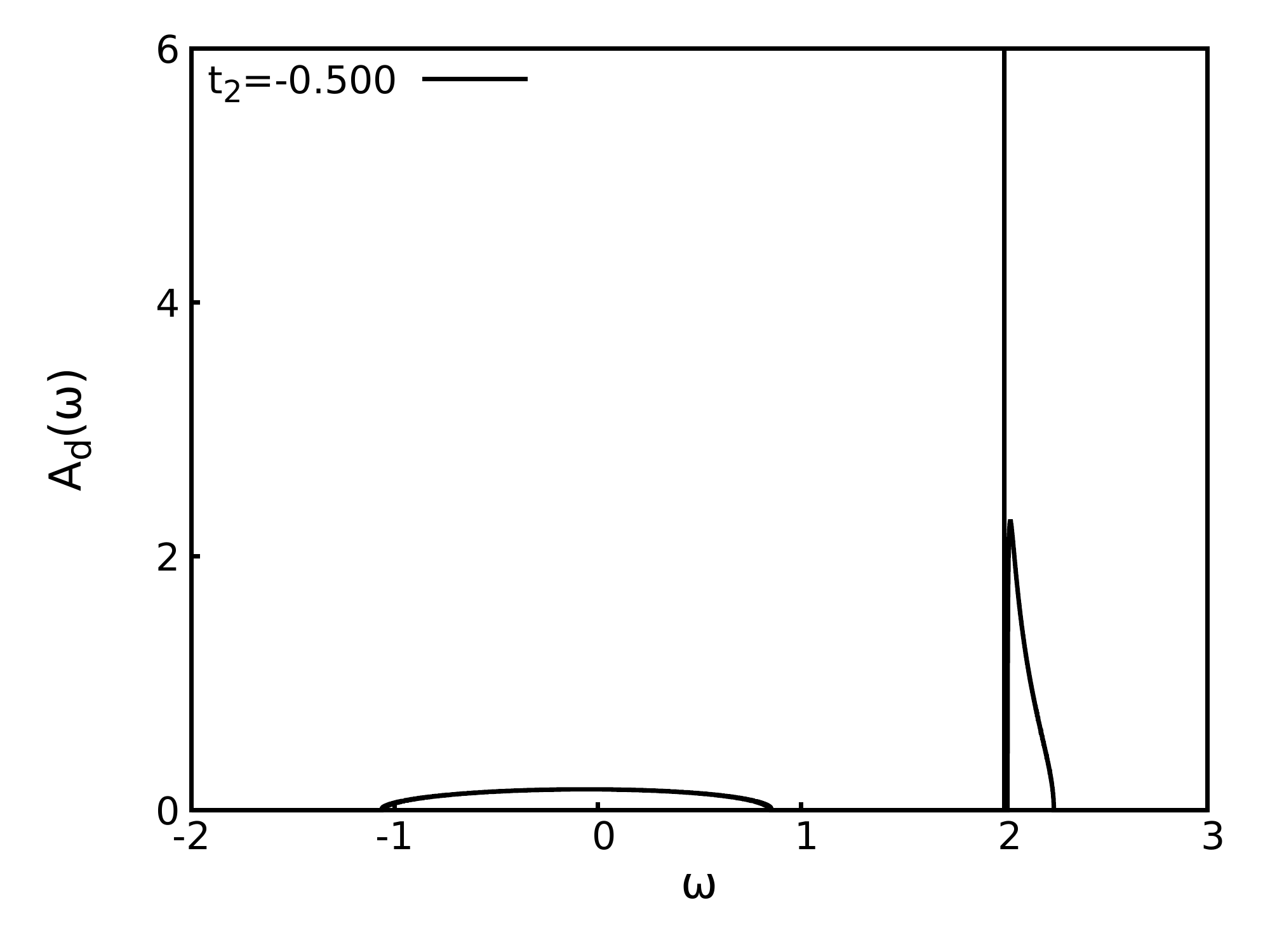}\qquad 
	\includegraphics[width=0.45\linewidth]{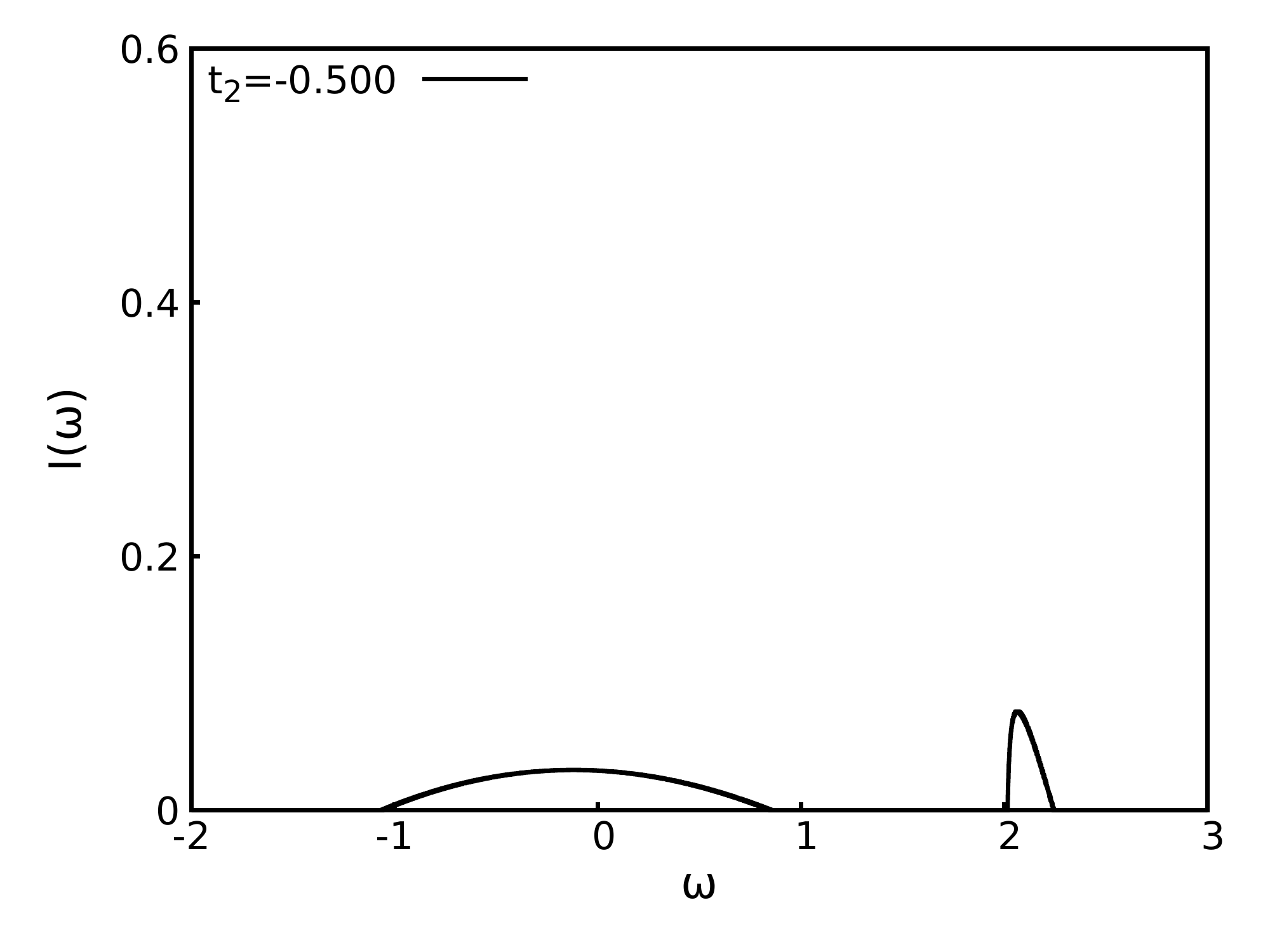}  
	\caption{Density of states \(A_d(\omega)\) and transport function \(I(\omega)\) for \(U=2.0\), \(n_f=0.75\), and \(t_2=-0.5\) (\(t^{++}=0\)).}
	\label{fig:dos-nf075t2-05}
\end{figure}


\begin{figure}[!b]
	\centering
	\includegraphics[width=0.45\linewidth]{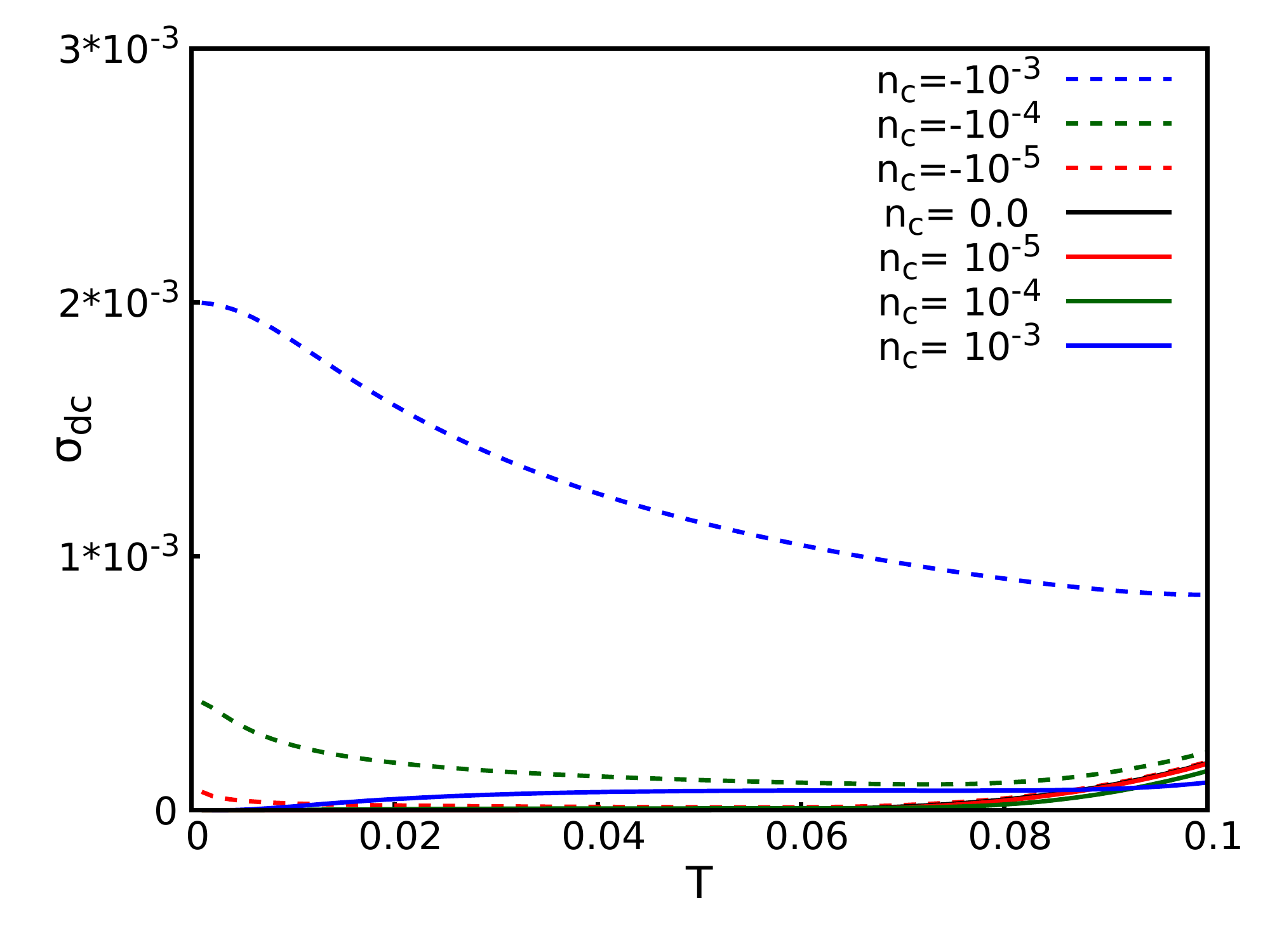} \qquad
	\includegraphics[width=0.45\linewidth]{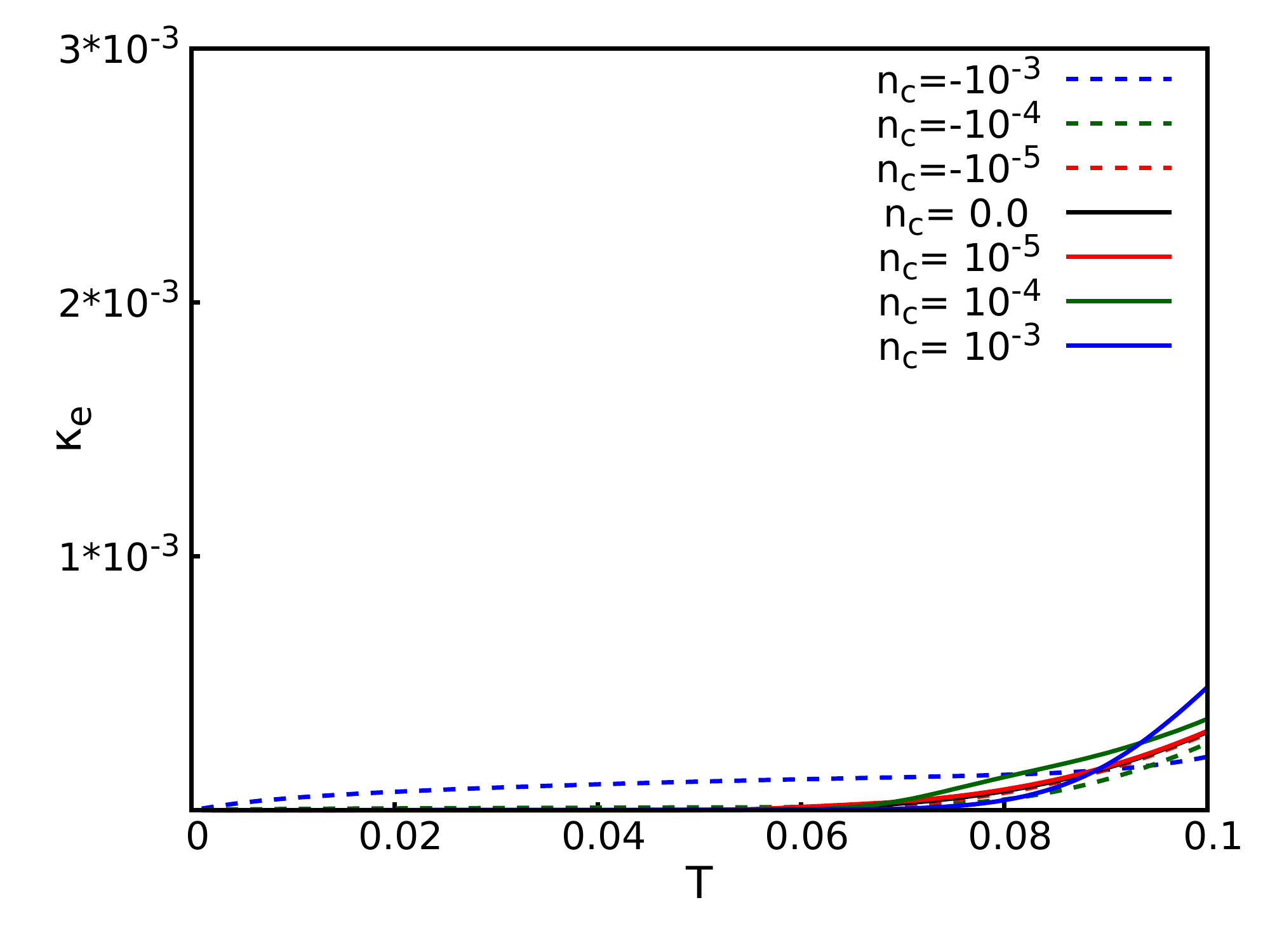} \\
	\includegraphics[width=0.45\linewidth]{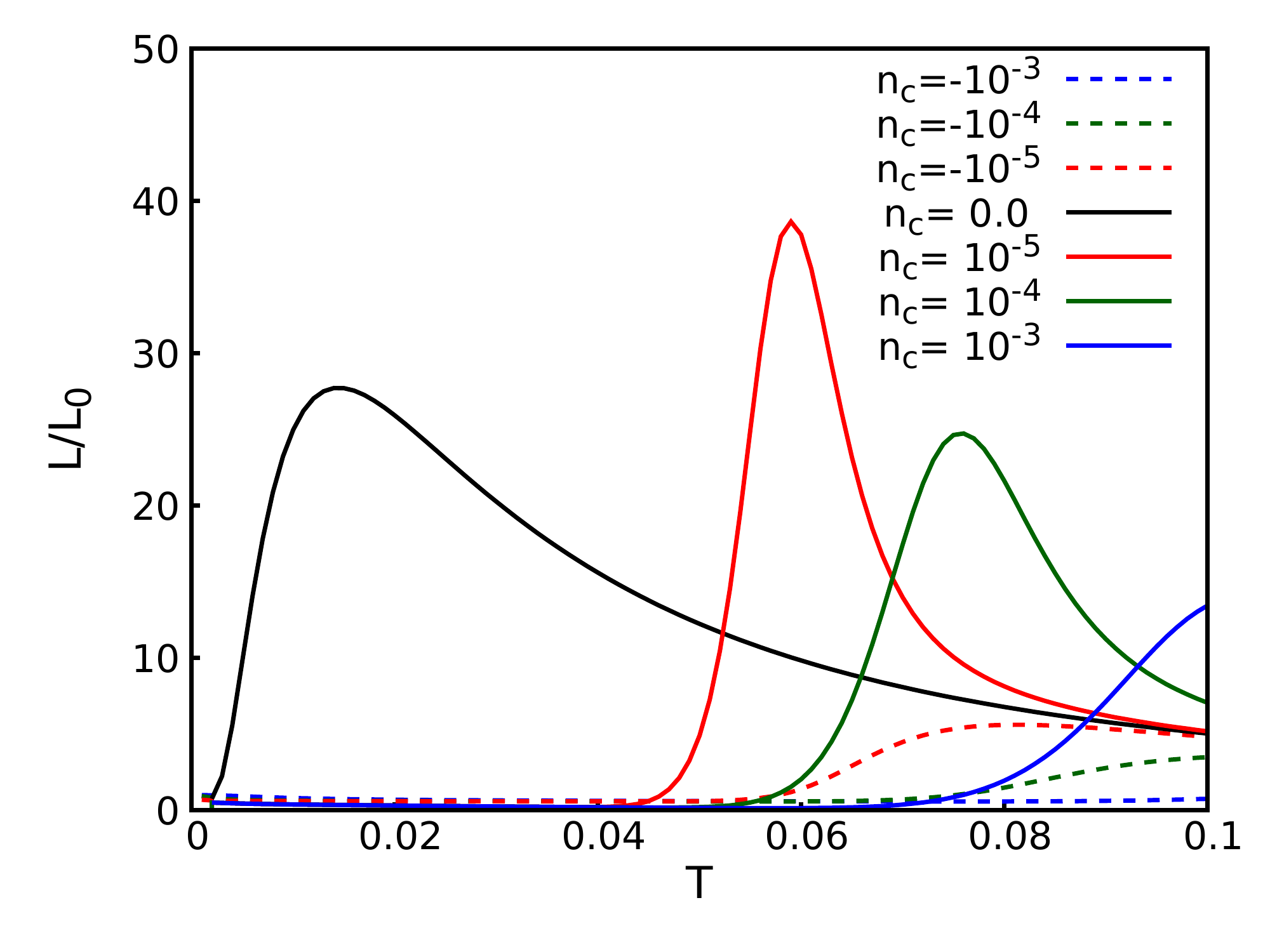} \qquad
	\includegraphics[width=0.45\linewidth]{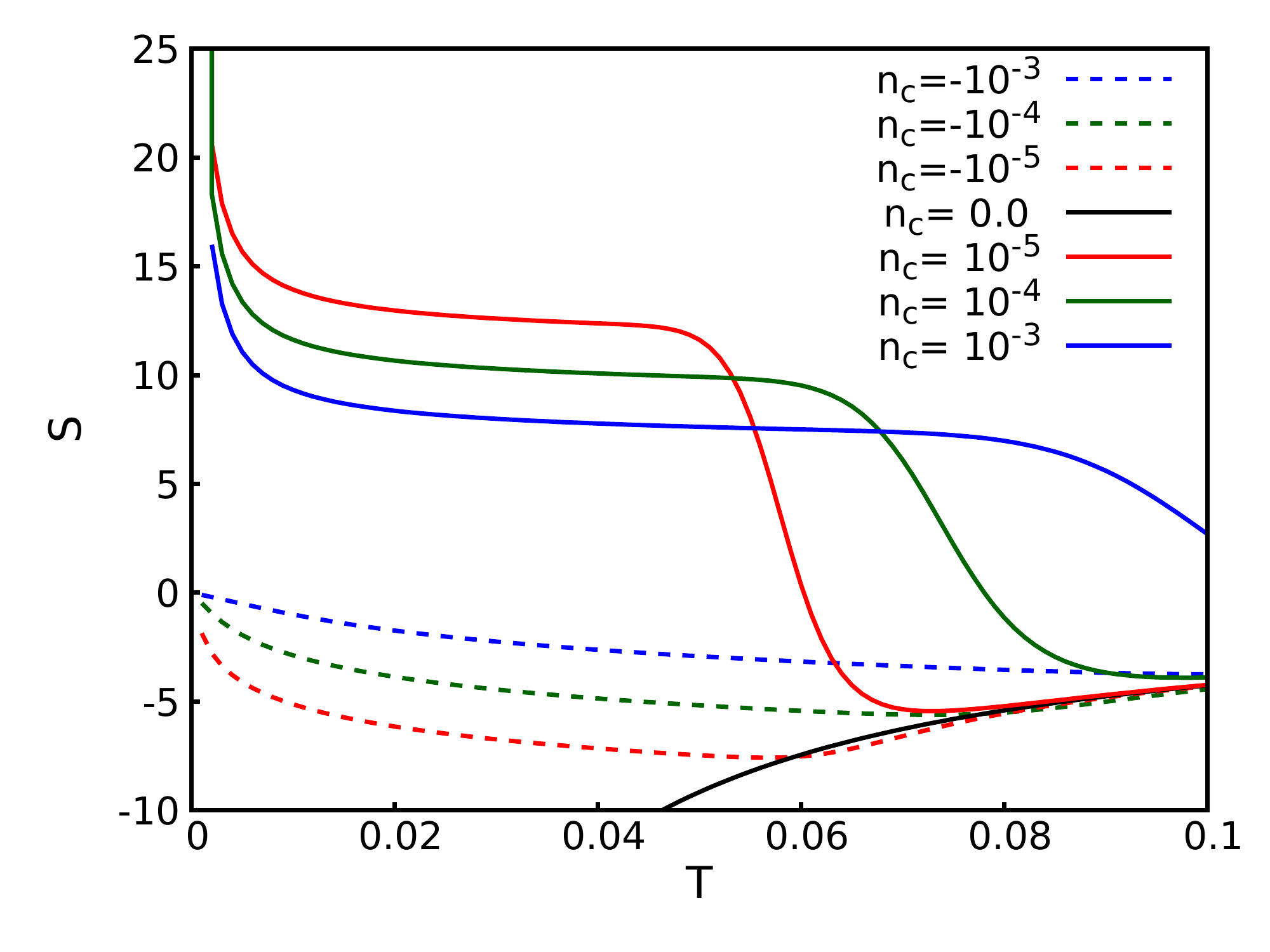} 
	\caption{(Colour online) Temperature dependences of the dc charge \(\sigma_{\textrm{dc}}\) and thermal \(\kappa_{\textrm{e}}\) conductivities, Lorenz number \(L\) (\(L_0=\piup^2/3\)), and thermopower (Seebeck coefficient) \(S\) for \(U=2.0\), \(n_f=0.75\), \(t_2=-0.5\) (\(t^{++}=0\)), \(n_d=0.25+n_c\).}
	\label{fig:therm-nf075t2-05nd025}
\end{figure}

Herein above we have considered the case of half-filling of \(f\) particle states, \(n_f=0.5\). However, in our previous investigations~\cite{dobushovskyi:125133}, it was found that for the case of \(f\) particle doping, \(n_f>0.5\), the third narrow band appears [figure~\ref{fig:nf075} (c) and (d)] when the hopping amplitude between the sites occupied by \(f\) particle becomes very small, \(|t^{++}|\ll t_1\). Now, the DOS contains three bands: the lower and the upper Hubbard bands with equal spectral weights \(w_0=1-n_f\) and the narrow middle band with spectral weight \(2n_f-1\) arising from the localized \(d\) electron states in the clusters of sites with occupied \(f\) states formed at \(n_f>0.5\). Hence, one can consider two cases of Mott insulators: the large gap Mott insulator for \(n_d=1-n_f\) and the small gap Mott insulator for \(n_d=n_f\). For \(t^{++}=0\) (\(t_2=-0.5\)), this middle band shrinks to a level with a \(\delta\)-peak on DOS, whereas it gives no contribution to the transport function (figure~\ref{fig:dos-nf075t2-05}). Hence, the gaps on the DOS and transport function are of different width. In the cases of a pure large gap insulator, \(n_d=1-n_f+n_c\) with \(n_c=0\), at very low temperatures, \(T\to0\), the chemical potential is placed in the centre of large gap, i.e.,  between the top of lower Hubbard band and \(\delta\)-peak from localized states on DOS. On the other hand, these localized states do not contribute to transport coefficients, and the transport function \(I(\omega)\) ``feels'' the larger gap between the top of the lower Hubbard band and the bottom of the upper one. However, now, the chemical potential shifts from the centre of the gap on the transport function leading to an additional strong enhancement of thermopower \(S(T)\) at \(T\to0\), not seen in figure~\ref{fig:therm-nf075t2-05nd025}, which causes a strong reduction of the Lorenz number by the second term in \eqref{eq:LN} at low temperatures, in contrast to the divergent behaviour observed for typical Mott insulators.
For the lightly hole doped large gap insulator, \(n_c<0\), the behaviour is similar to the one considered above (figure~\ref{fig:therm-nf075t2-05nd025}). On the other hand, for the case of electron doping, \(n_c>0\), and at low temperatures, the chemical potential is placed in the middle level of localized states separated by a small gap from the upper Hubbard band which it never enters. Now, the transport function is always very asymmetric on different sides of the Fermi level, within the Fermi window, producing large enhancement of the Seebeck coefficient at low temperatures. For the small gap insulator at \(n_d=n_f+n_c\), we observe almost metallic transport at high temperatures and an enhancement of thermopower at very low temperatures. 

\begin{figure}[!t]
	\centering
	\includegraphics[width=0.45\linewidth]{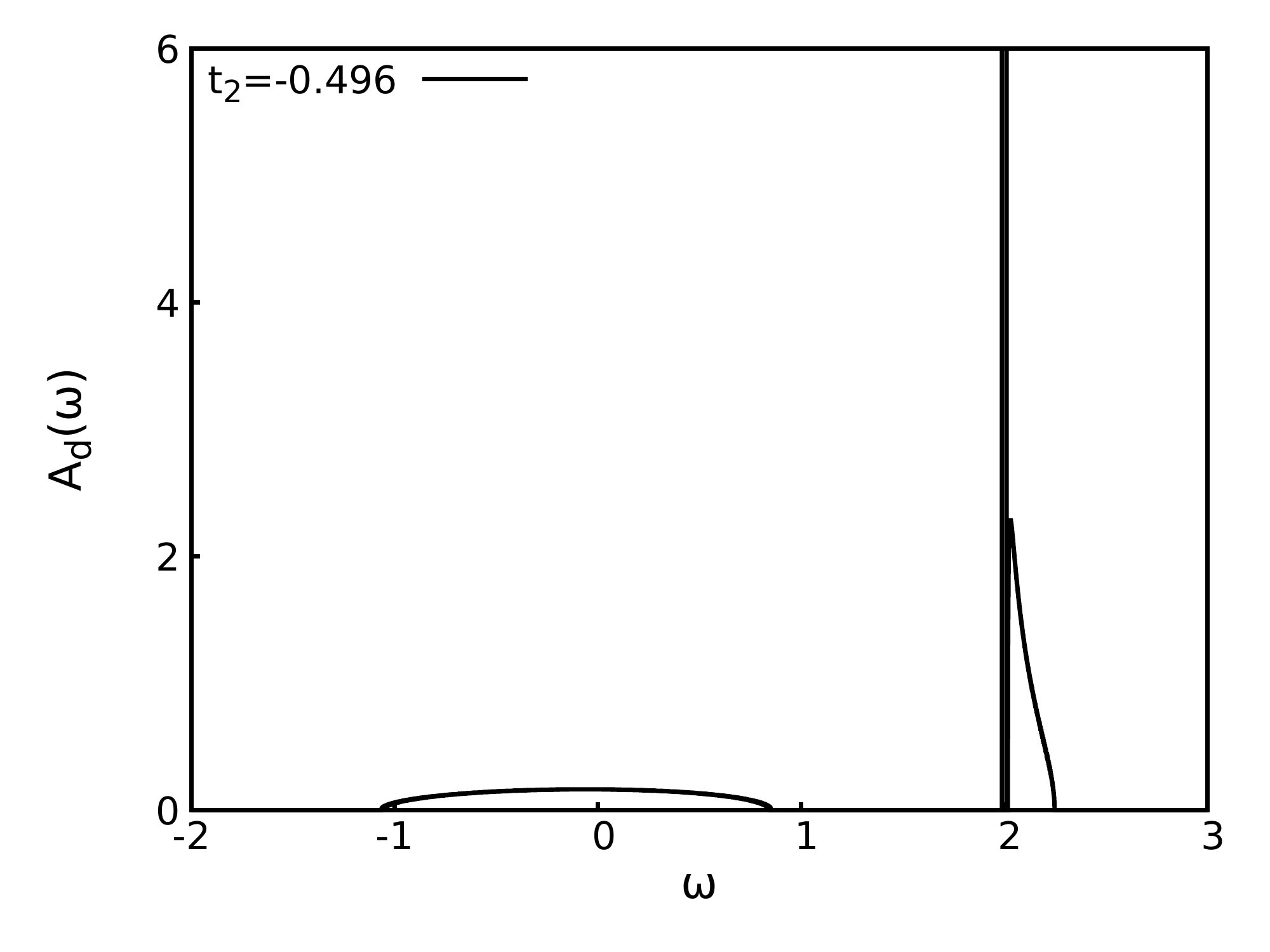}\qquad
	\includegraphics[width=0.45\linewidth]{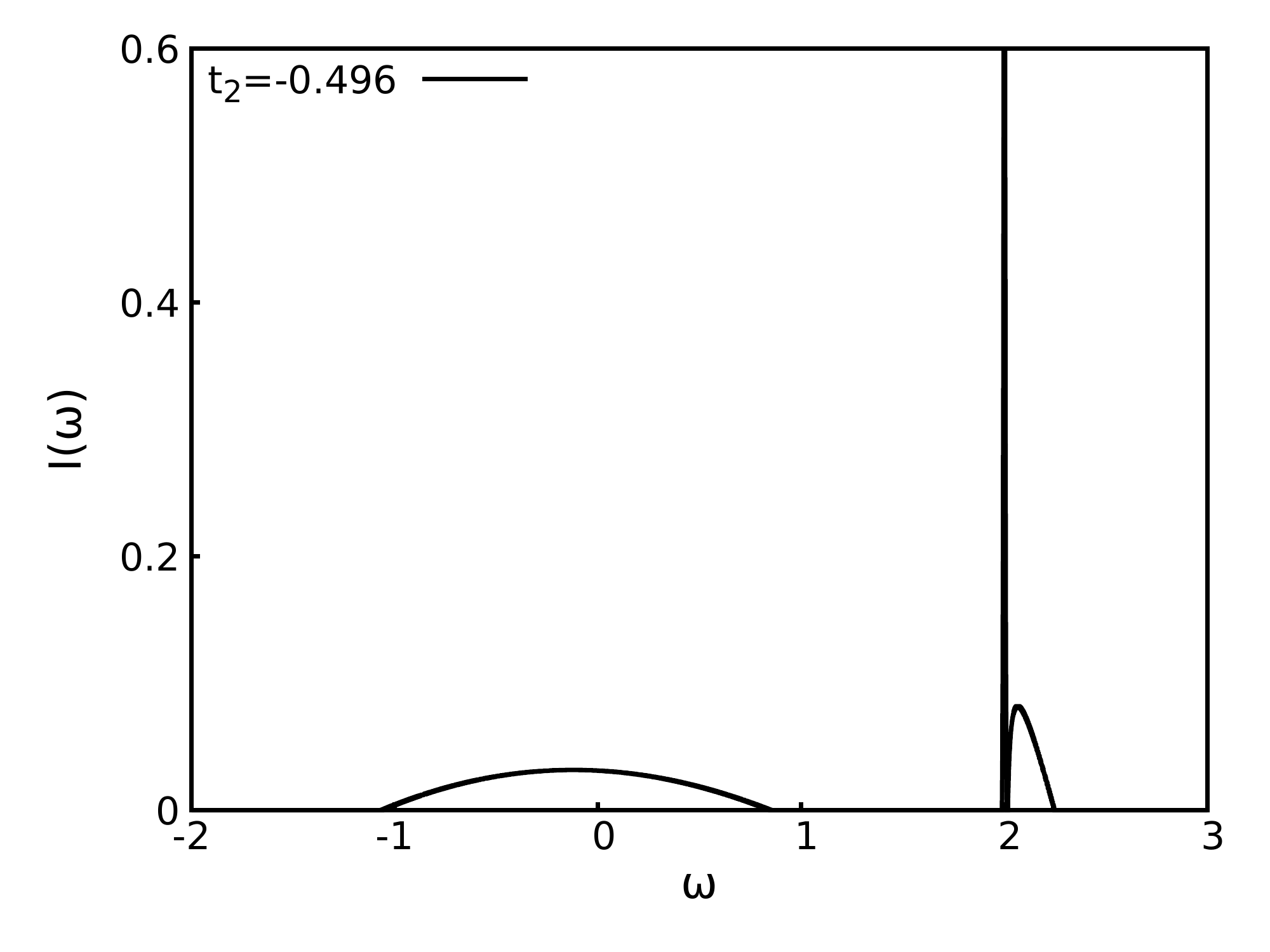} 
	\caption{Density of states \(A_d(\omega)\) and transport function \(I(\omega)\) for \(U=2.0\), \(n_f=0.75\), and \(t_2=-0.496\) (\(t^{++}=0.008\)).}
	\label{fig:dos-nf075t2-0496}
\end{figure}


\begin{figure}[!b]
	\centering
	\includegraphics[width=0.45\linewidth]{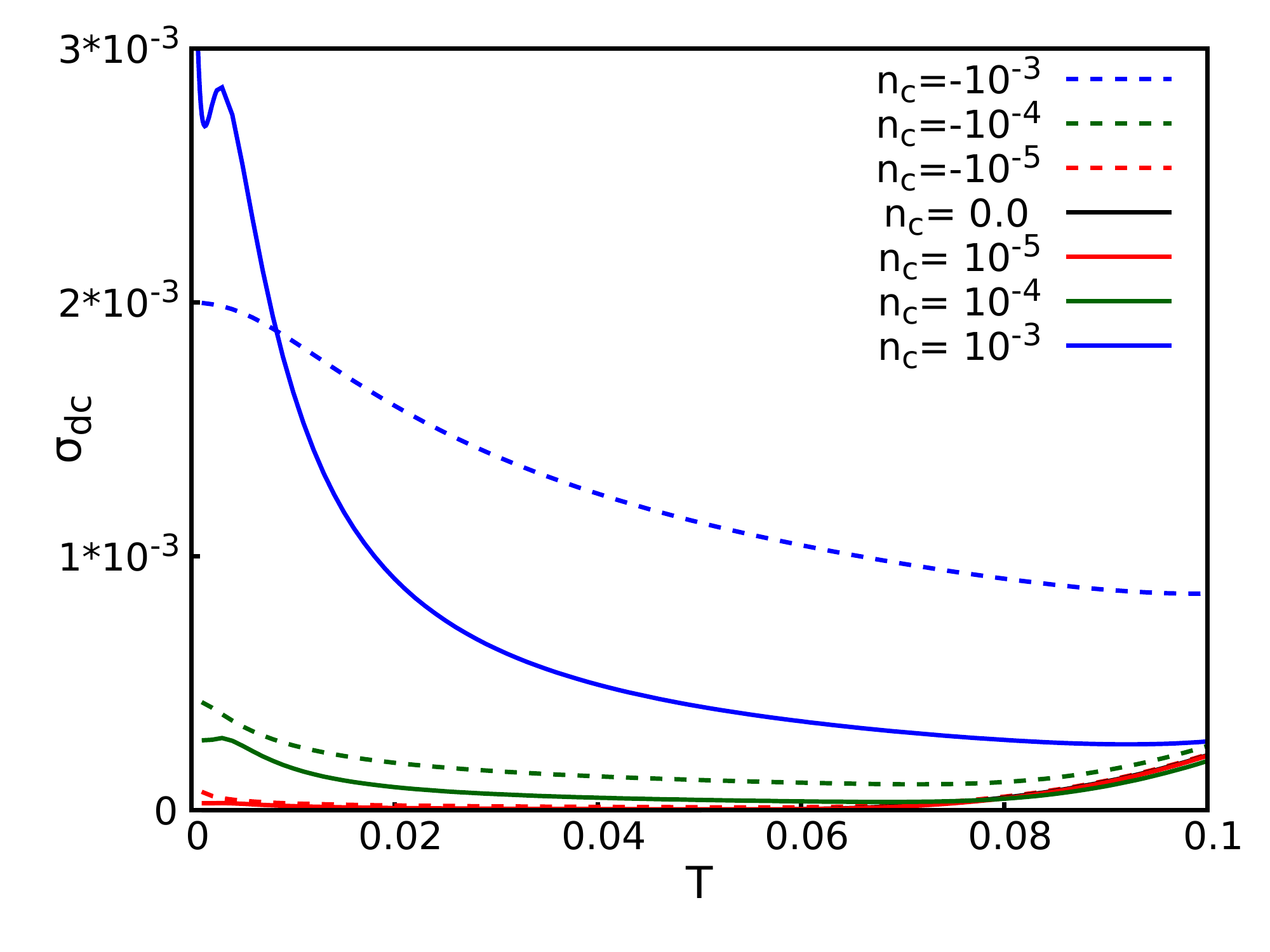} \qquad
	\includegraphics[width=0.45\linewidth]{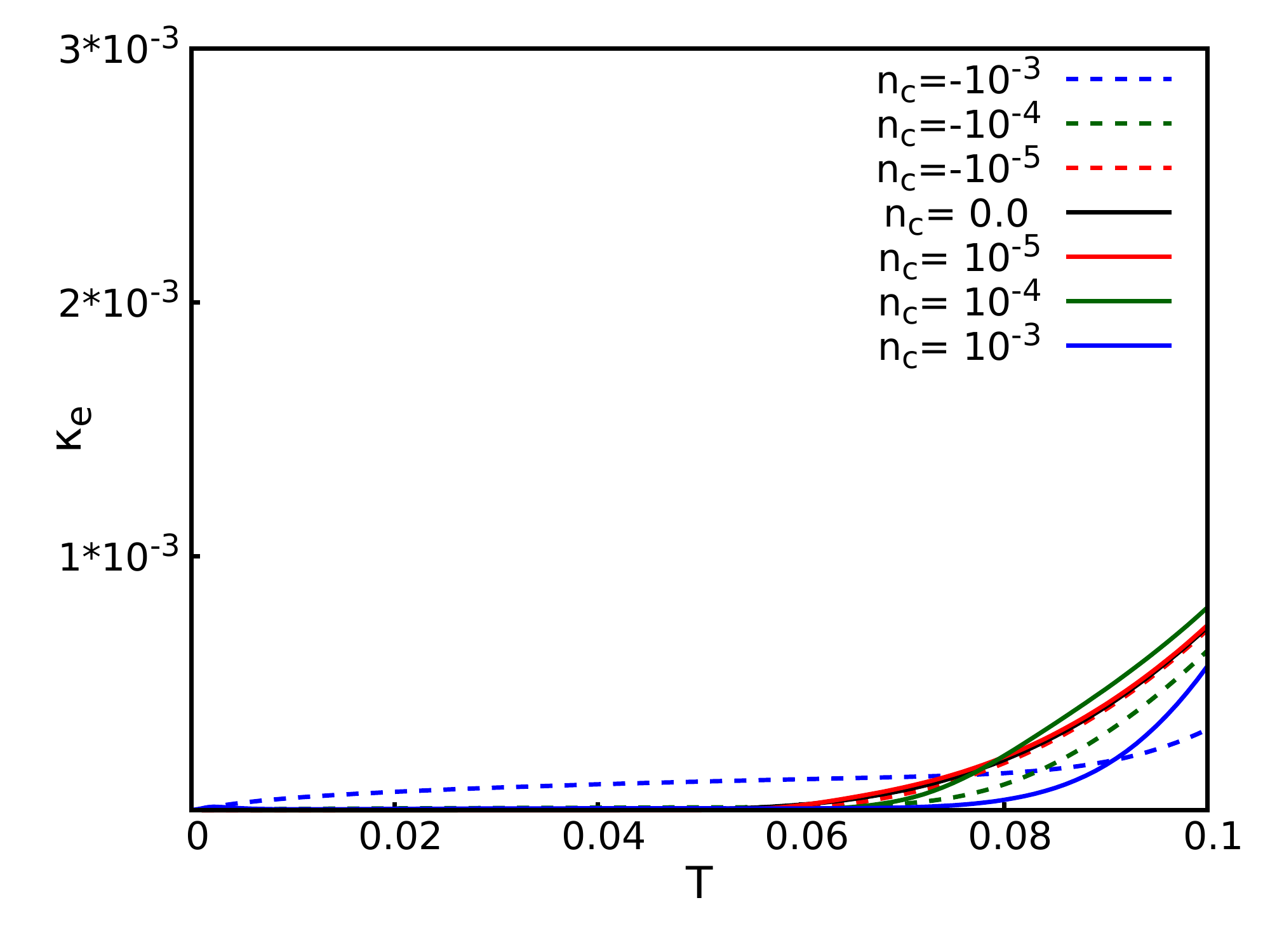} \\
	\includegraphics[width=0.45\linewidth]{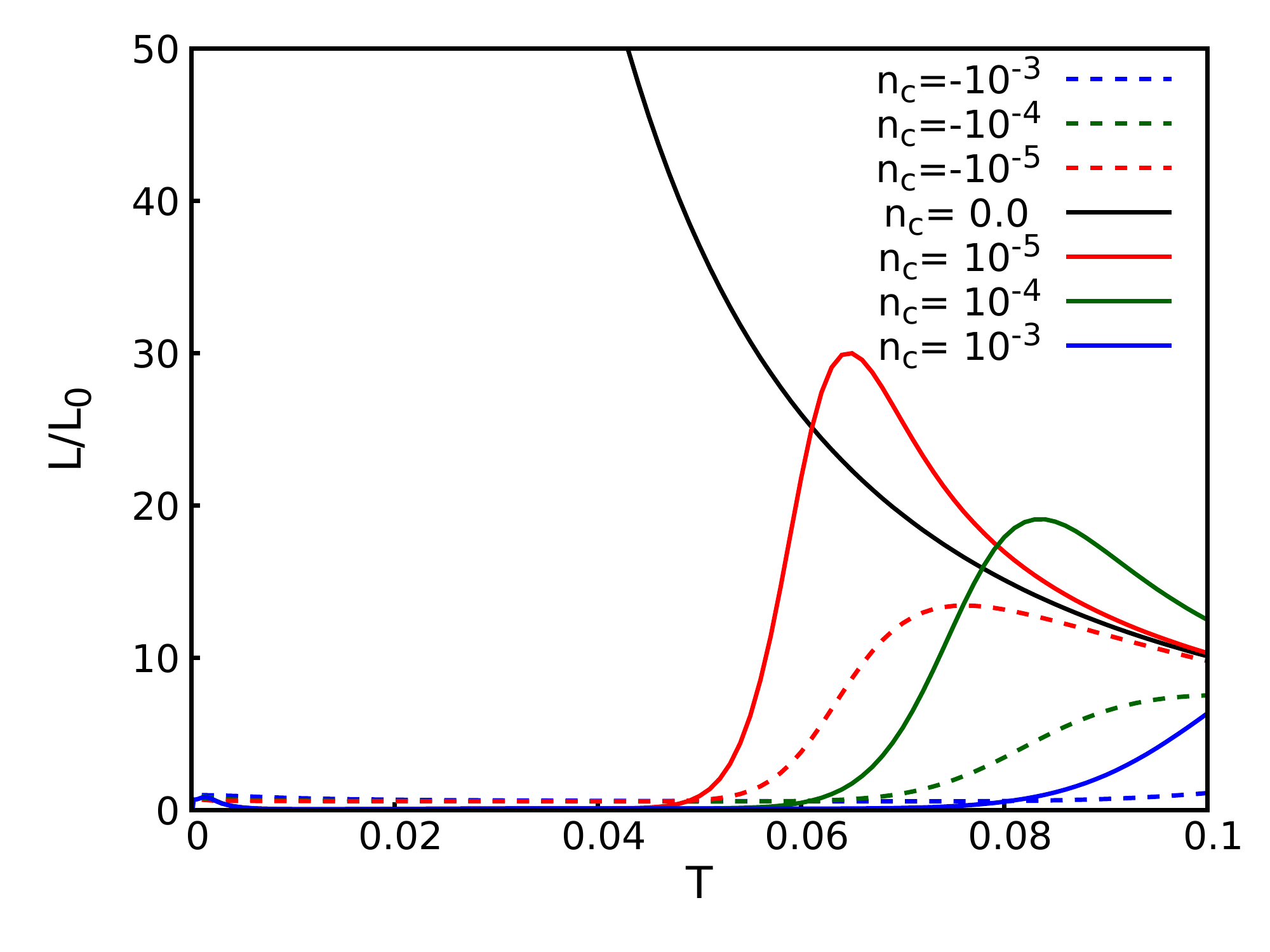} \qquad
	\includegraphics[width=0.45\linewidth]{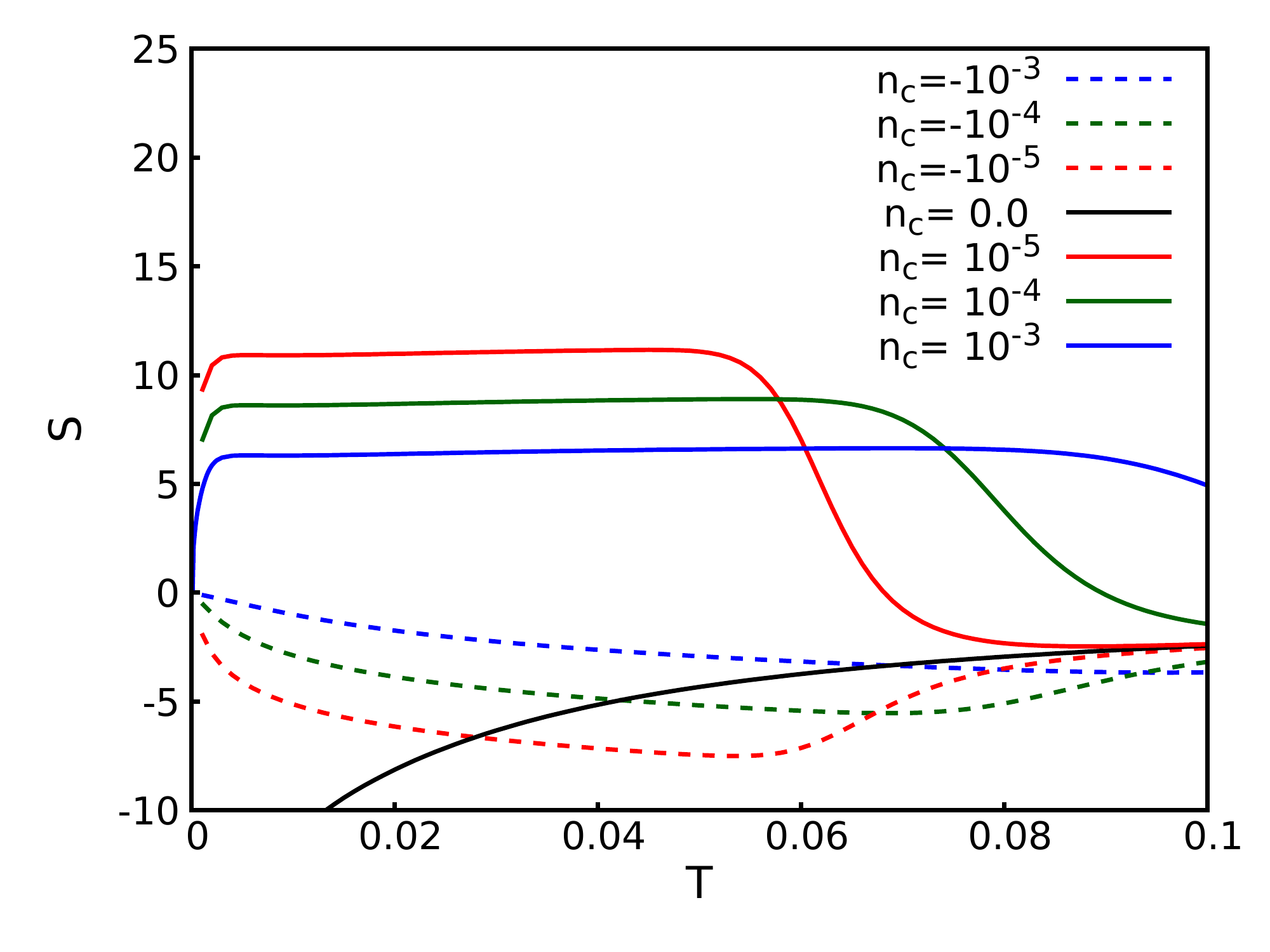} 
	\caption{(Colour online) Temperature dependences of the dc charge \(\sigma_{\textrm{dc}}\) and thermal \(\kappa_{\textrm{e}}\) conductivities, Lorenz number \(L\) (\(L_0=\piup^2/3\)), and thermopower (Seebeck coefficient) \(S\) for \(U=2.0\), \(n_f=0.75\), \(t_2=-0.496\) (\(t^{++}=0.008\)), \(n_d=0.25+n_c\).} 
	\label{fig:therm-nf075t2-0496nd025}
\end{figure}

The switching on of the hopping between the sites occupied by \(f\) electrons, \(t^{++}>0\) (\(t_2<-0.5\)), leads to the spreading of the \(\delta\)-peak from localized states on DOS into the narrow band, and the narrow resonant peak appears on the transport function, figure~\ref{fig:dos-nf075t2-0496}. Now, the gaps on the DOS and transport function are of the same width, which restores the typical behaviour for the Mott insulator case. The transport properties, figure~\ref{fig:therm-nf075t2-0496nd025}, of the hole doped Mott insulator, \(n_d=1-n_f+n_c\) with \(n_c<0\), are similar to the one discussed in the previous case of \(t^{++}=0\), but one can notice a strong enhancement of the dc charge conductivity for the electron doping, \(n_c>0\), when chemical potential enters the resonant peak on the transport function. Moreover, for electron doping, the Seebeck coefficient displays almost flat temperature dependence in a wide temperature range and it starts to decrease in an expected manner at \(T\to0\).

\begin{figure}[!b]
	\centering
	\includegraphics[width=0.45\linewidth]{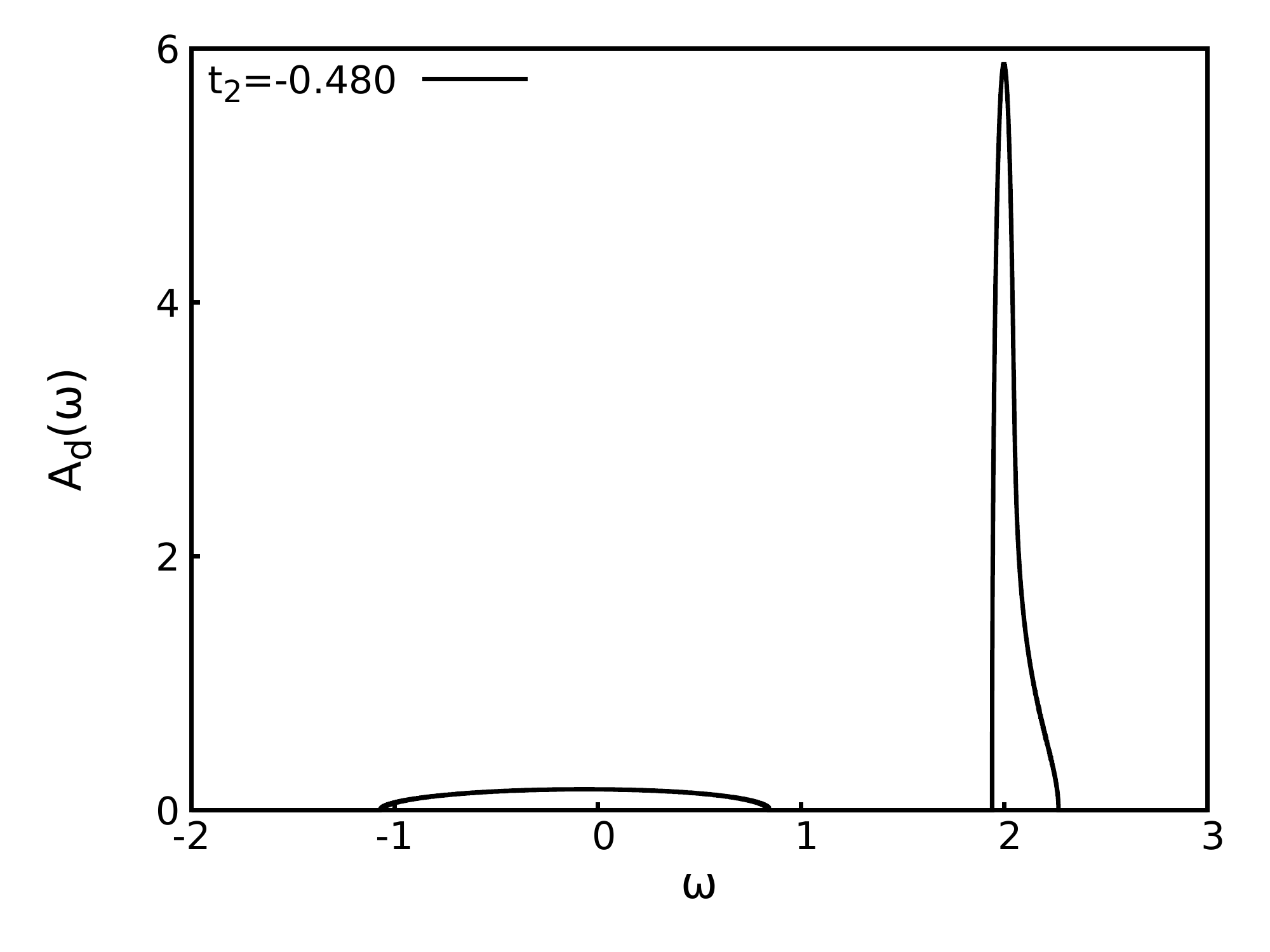} \qquad
	\includegraphics[width=0.45\linewidth]{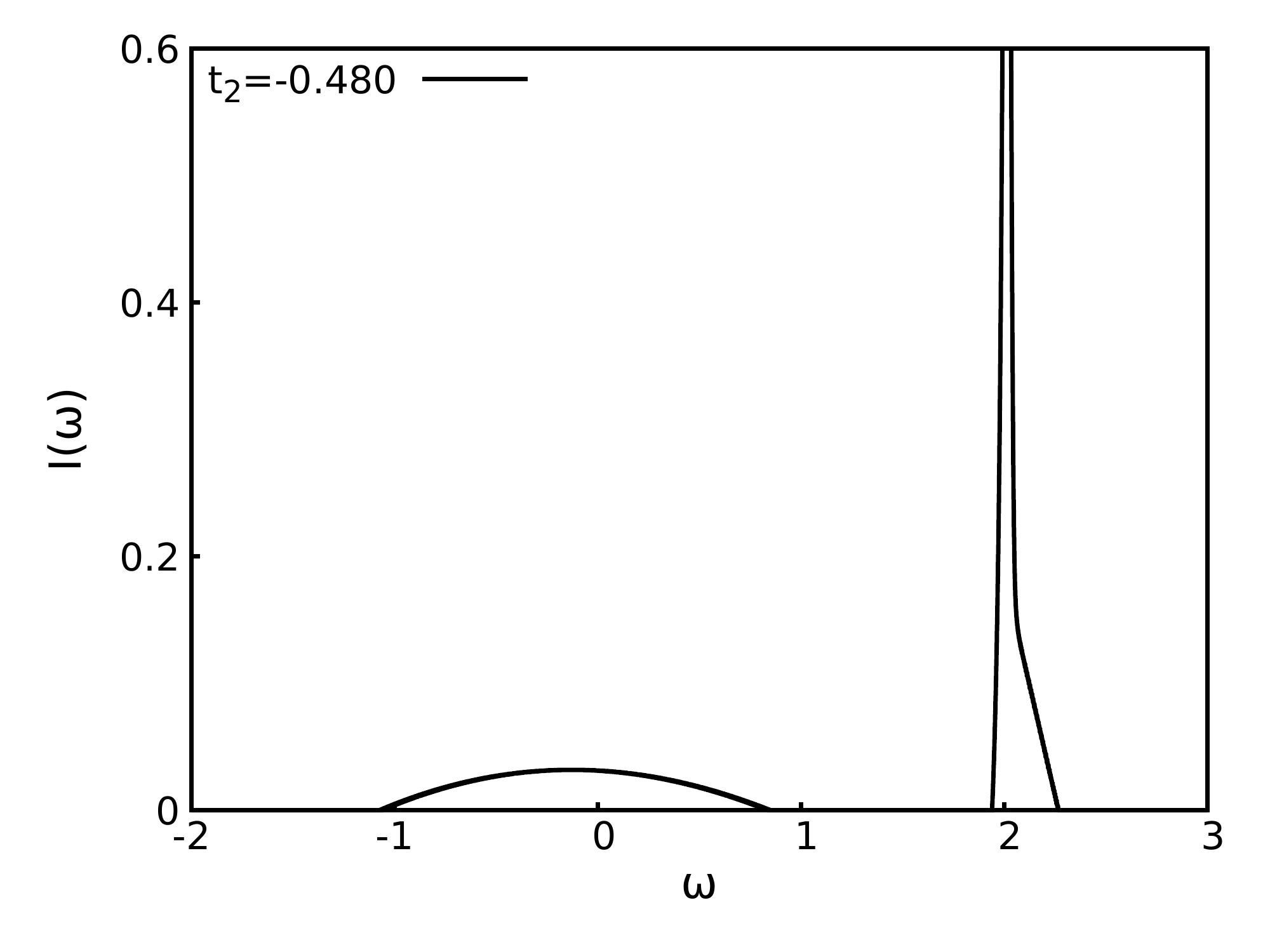} 
	\caption{Density of states \(A_d(\omega)\) and transport function \(I(\omega)\) for \(U=2.0\), \(n_f=0.75\), and \(t_2=-0.48\) (\(t^{++}=0.04\)).}
	\label{fig:dos-nf075t2-048}
\end{figure}


\begin{figure}[!b]
	\centering
	\includegraphics[width=0.45\linewidth]{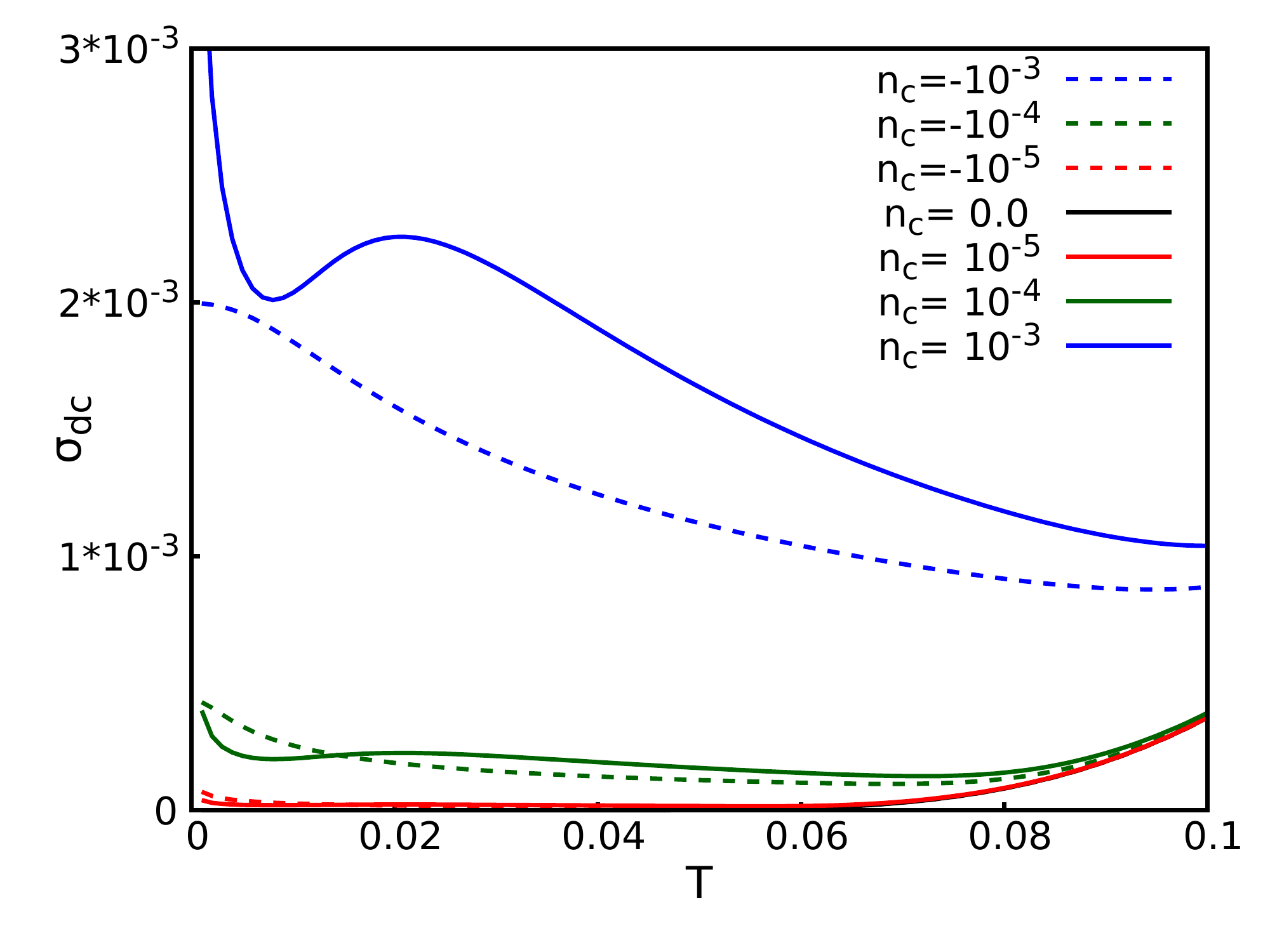} \qquad
	\includegraphics[width=0.45\linewidth]{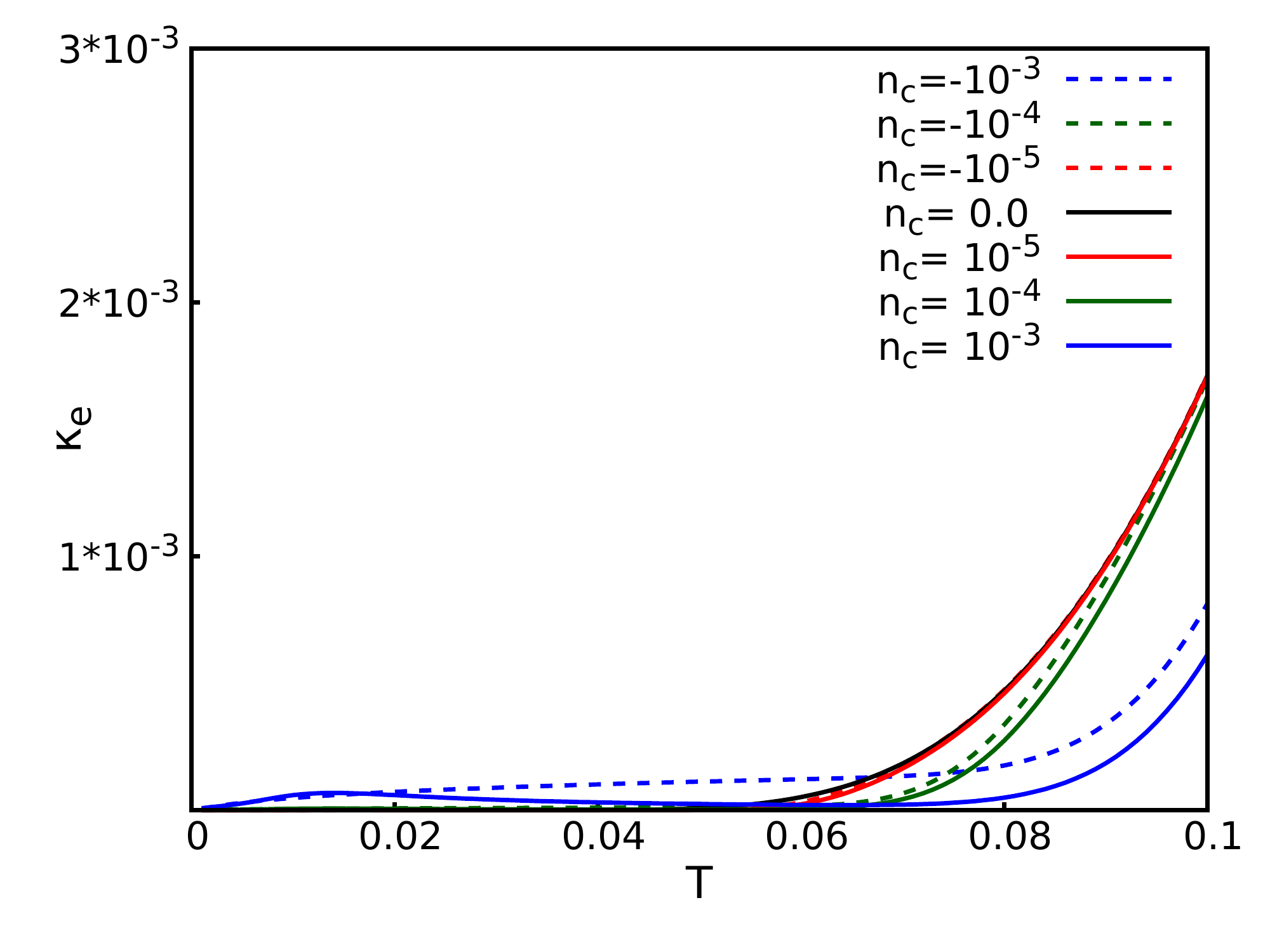} \\
	\includegraphics[width=0.45\linewidth]{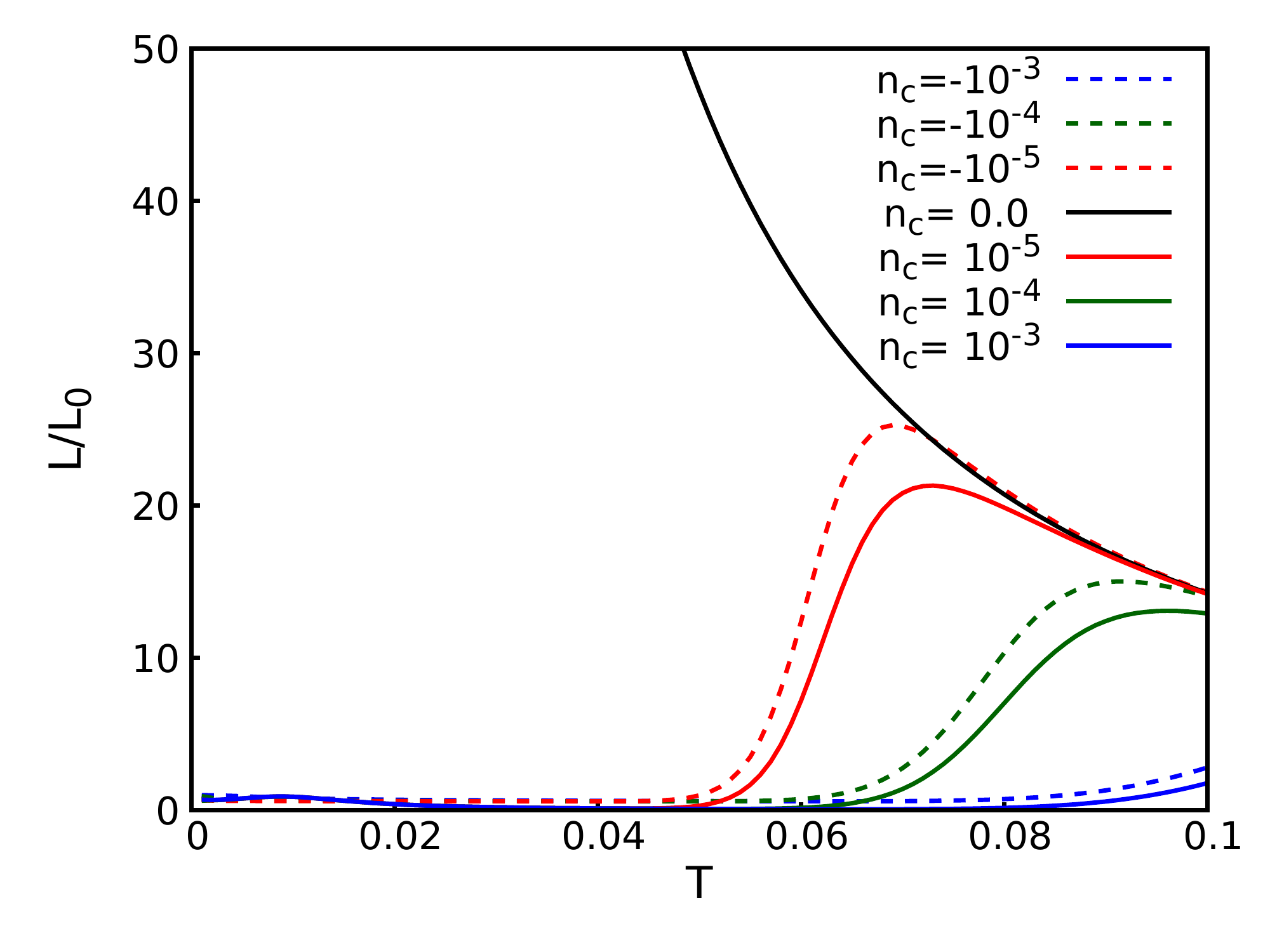} \qquad
	\includegraphics[width=0.45\linewidth]{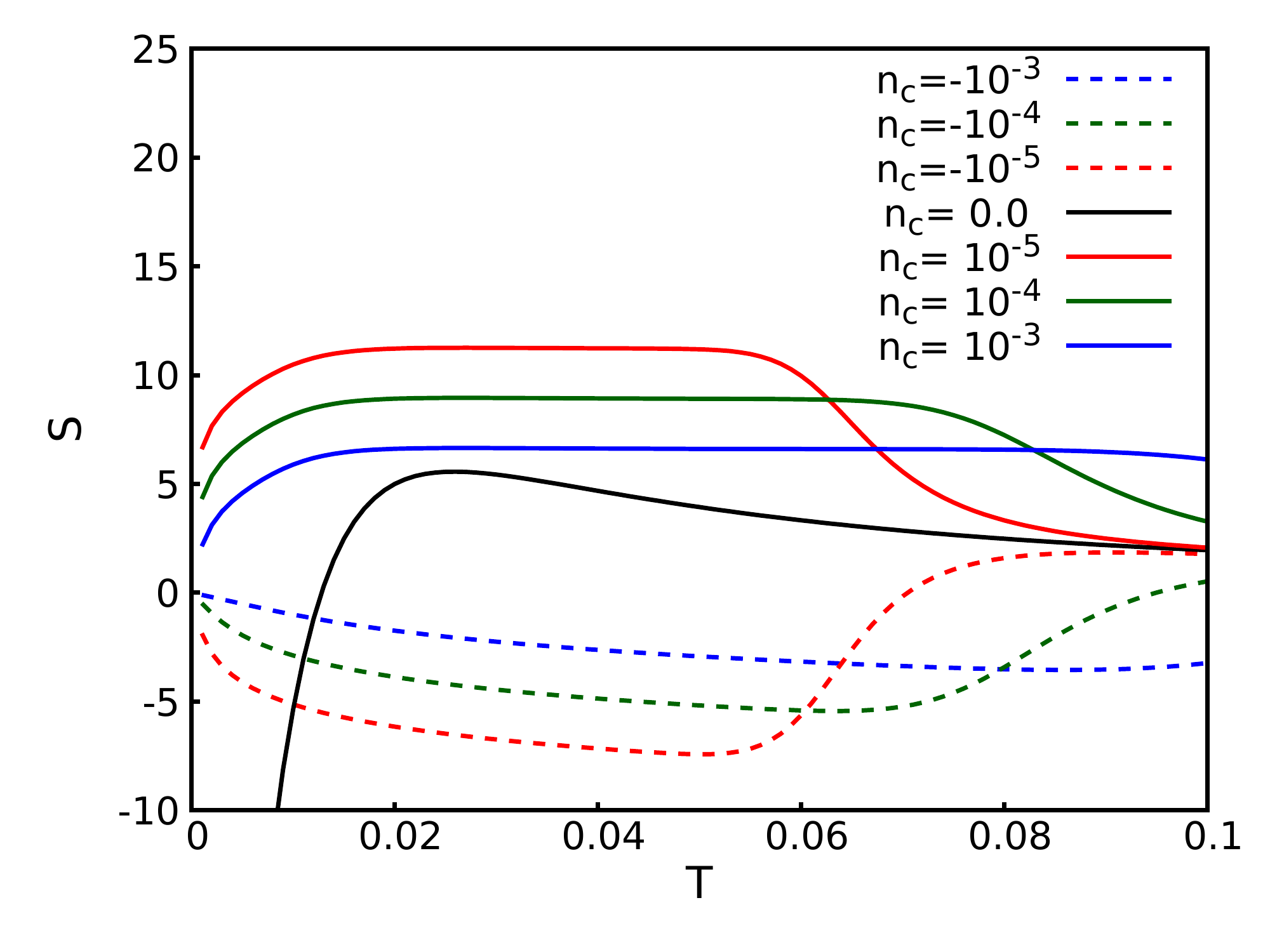} 
	\caption{(Colour online) Temperature dependences of the dc charge \(\sigma_{\textrm{dc}}\) and thermal \(\kappa_{\textrm{e}}\) conductivities, Lorenz number \(L\) (\(L_0=\piup^2/3\)), and thermopower (Seebeck coefficient) \(S\) for \(U=2.0\), \(n_f=0.75\), \(t_2=-0.48\) (\(t^{++}=0.04\)), \(n_d=0.25+n_c\).} 
	\label{fig:therm-nf075t2-048nd025}
\end{figure}

For larger values of the hopping amplitude \(t^{++}\) between the sites occupied by \(f\) electrons, the middle band of localized states joins with the upper Hubbard band. Now, we observe a huge enhancement at the bottom of the upper Hubbard band both on DOS and on the transport function due to resonant peak, figure~\ref{fig:dos-nf075t2-048}. The temperature dependences of the dc charge and thermal conductivities are similar to the above discussed but the thermopower displays an anomaly for the case of Mott insulator, figure~\ref{fig:therm-nf075t2-048nd025}. Now, the Seebeck coefficient \(S(T)\) is positive  at high temperatures and increases with its decreasing up to some temperature value. For lower temperatures it starts to decrease, changes its sign and rapidly increases at low temperatures.

\section{Conclusions}\label{sec:conclusions}

In this article we have discussed the peculiarities of the charge and thermal transport in the Falicov-Kimball model with correlated hopping at a light doping of the Mott insulator phase. We consider the cases of the strongly reduced hopping amplitude \(t^{++}\) between the sites occupied by the \(f\) electrons, when the DOS and transport function display anomalous features, including edge singularity, resonant peak, and additional band of localized states at \(n_f>0.5\). At half-filling \(n_f=0.5\) and in the Mott insulator phase, \(n_c=0\), the dc charge and thermal conductivities display a typical behaviour for the large gap insulators with asymmetric DOS, whereas the light hole and electron doping restore the bad metallic conductivity with an enhanced thermopower for the electron doping case. Outside the half-filling case, when \(n_f>0.5\), and for the completely reduced hopping between the sites occupied by \(f\)-electrons, the gaps on the DOS and transport function do not coincide, which causes an anomalous thermoelectric transport at low temperatures featuring a strong reduction of the Lorenz number and a huge enhancement of thermopower for the electron doping case.


\ukrainianpart

\title{Термоелектричні властивості діелектрика Мотта з корельованим переносом при мікролегуванні}
\author{Д.А. Добушовський, А.М. Швайка}
\address{
	Інститут фізики конденсованих систем НАН України, вул. І. Свєнціцького, 1, 79011 Львів, Україна
}

\makeukrtitle

\begin{abstract}
	\tolerance=3000%
	Обговорюється вплив індукованої корельованим переносом локалізації колективізованих електронів на електронний транспорт заряду і тепла в слабко легованій фазі моттівського діелектрика моделі Фалікова-Кімбала. Детально розглядається випадок сильно редукованого переносу між вузлами із заповненими рівнями \(f\)-електронів, коли на густині станів виникає додаткова зона локалізованих станів \(d\)-електронів у моттівській щілині. Внаслідок сильної електрон-діркової асиметрії і появи аномальних особливостей на густині станів та транспортній функції, спостерігається сильне зростання коефіцієнта Зеєбека при низьких температурах, коли він слабко змінюється в широкому діапазоні температур.
	\keywords термоелектричні явища, моттівські діелектрики, локалізація, модель Фалікова-Кімбала, корельований перенос, теорія динамічного середнього поля
	
\end{abstract}

\end{document}